\let\mypdfximage\pdfximage
\def\pdfximage{\immediate\mypdfximage}
\documentclass[useAMS,usenatbib]{mn2e}
\usepackage{color} 
\usepackage[pdftex]{graphicx}
\usepackage[maxfloats=256]{morefloats}
\usepackage{multirow}
\usepackage{amssymb}

\usepackage{amsmath}

\newcommand{\kpc}{\mbox{${\rm kpc}$}} 
\newcommand{\pc}{\mbox{${\rm pc}$}} 
\newcommand{\kms}{\mbox{${\rm km\, s^{-1}}$}}
\newcommand{\kmskpc}{\mbox{${\kpc\, \kms}$}}

\newcommand{\Myr}{\mbox{$\>{\rm Myr}$}}
\newcommand{\Gyr}{\mbox{$\>{\rm Gyr}$}}

\newcommand{\Msun}{\>{\rm M_{\odot}}}
\newcommand{\Rd}{\mbox{$R_{\rm d}$}} 
\newcommand{\zd}{\mbox{$z_{\rm d}$}} 
\newcommand{\md}{\mbox{$M_{\rm d}$}}

\newcommand{\avg}[1]{\mbox{$\left<{#1}\right>$}}
\newcommand{\sig}[1]{\mbox{$\sigma_{#1}$}}

\newcommand{\feh}{\mbox{$\rm [Fe/H]$}}

\newcommand{\act}[1]{\mbox{$\rm J_{#1,0}$}}
\newcommand{\ac}[1]{\mbox{$\rm J_{#1}$}}

\newcommand{\omr}{\mbox{$\Omega_{R,0}$}}
\newcommand{\omz}{\mbox{$\Omega_{z,0}$}}

\def\etc{{\it etc.}}
\def\ie{{\it i.e.}}

\long\def\Ignore#1{\relax}

\title[B/P bulges in action space]{Box/peanut-shaped bulges in action space}
   
\author[Debattista et al.]{Victor P. Debattista$^1$\thanks{E-mail: 
vpdebattista@gmail.com}, David J. Liddicott$^1$, Tigran
Khachaturyants$^1$, \newauthor Leandro Beraldo e Silva$^1$ \\
$^{1}$ Jeremiah Horrocks Institute, University of Central
  Lancashire, Preston, PR1 2HE, UK \\
}

\begin{document}   

\date{{\it Draft version on \today}}
\pagerange{\pageref{firstpage}--\pageref{lastpage}} \pubyear{----}
\maketitle

\label{firstpage}

\begin{abstract}
We introduce the study of box/peanut (B/P) bulges in the action space
of the initial axisymmetric system.  We explore where populations with
different actions end up once a bar forms and a B/P bulge develops.
We find that the density bimodality due to the B/P bulge (the X-shape)
is better traced by populations with low radial, \act{R}, or vertical,
\act{z}, actions, or high azimuthal action, \act{\phi}.  Generally
populations separated by \act{R}\ have a greater variation in bar
strength and vertical heating than those separated by \act{z}.  While
the bar substantially weakens the initial vertical gradient of
\act{z}, it also drives a strikingly monotonic \emph{vertical} profile
of \act{R}.
We then use these results to guide us in assigning metallicity to star
particles in a pure $N$-body model.  Because stellar metallicity in
unbarred galaxies depends on age as well as radial and vertical
positions, the initial actions are particularly well suited for
assigning metallicities.  We argue that assigning metallicities based
on single actions, or on positions, results in metallicity
distributions inconsistent with those observed in real galaxies.  We
therefore use all three actions to assign metallicity to an $N$-body
model by comparing with the actions of a star-forming, unbarred
simulation.  The resulting metallicity distribution is pinched on the
vertical axis, has a realistic vertical gradient and has a stronger
X-shape in metal-rich populations, as found in real galaxies.
\end{abstract}

\begin{keywords}
  Galaxy: abundances --- Galaxy: bulge --- Galaxy: evolution ---
  Galaxy: stellar content --- Galaxy: structure --- galaxies: bulges
\end{keywords}

%

\section{Introduction}
\label{s:intro}

Among the integrals of motion that have been used to study the Milky
Way (MW), actions have found particular favour in part because they
can be computed directly from stellar orbits
\citep{binney_tremaine08}, or approximated from instantaneous positions 
and velocities under the assumption that the potential is locally of
St\"ackel form \citep{binney12}.  The adiabatic growth of the MW's
mass conserves the actions and stars should retain some memory of the
state of the MW when the stars were born.  However various resonant
processes alter the actions.  Scattering, such as by giant molecular
clouds, drives a slow diffusion in action space
\citep{binney_lacey88} which manifests as the slow heating of stellar
populations \citep{spitzer_schwarzschild51, spitzer_schwarzschild53,
wielen77, lacey84, icke82, villumsen85}.  Scattering can be viewed as
the resonant response to collective modes excited by fluctuations
\citep{nelson_tremaine99, heyvaerts10, sellwood13, sellwood15,
heyvaerts+17, fouvry+18}.  Actions are changed more substantially at
resonances of global perturbations, such as the inner and outer
Lindblad resonances \citep{lynden-bell_kalnajs72, barbanis_woltjer67,
carlberg_sellwood85}, and the corotation resonance
\citep{sellwood_binney02, roskar+10, daniel_wyse15} of bars and
spirals.  The resulting diffusion in action space can be anisotropic;
for instance during the migration driven by the spiral corotation
resonance, the radial action is largely conserved, while the azimuthal
action changes dramatically \citep{sellwood_binney02, roskar+10}. The
vertical action also is not conserved during corotation migration
because of vertical resonances \citep{solway+12, vera-ciro_donghia16}.

During dynamical instabilities the potential can change rapidly.  In
particular, the bar \citep{toomre81, sellwood81} and the buckling
instabilities \citep{raha+91} are generally violent and unlikely to
conserve the actions.  Notwithstanding this, the initial actions
determine how the stars react to the perturbation and thus help in
understanding the final distributions of stars within a B/P bulge once
it forms.  \citet{debattista+17} showed that the initial radial random
motions play a large role in determining the morphology of different
populations within the B/P bulge.  The formation of B/P bulges has a
long history of study using numerical simulations
\citep{combes_sanders81, combes+90, pfenniger_friedli91, raha+91,
merritt_sellwood94, berentzen+98, oneill_dubinski03,
martinez-valpuesta_shlosman04, debattista+04, debattista+05,
debattista+06, athanassoula05, bureau_athanassoula05,
martinez-valpuesta+06, jshen+10, martinez-valpuesta_gerhard11,
martinez-valpuesta_gerhard13, saha_gerhard13, saha+13, li_shen12,
li_shen15, pdimatteo+14, pdimatteo+15, pdimatteo16, fragkoudi+17,
fragkoudi+18, smirnov_sotnikova19}.  In this paper we introduce the
study of B/P bulges in the space of the initial actions.  This leads
to new insights into how B/P bulges form and suggests new avenues of
study.

A further motivation for exploring how initial actions map into a B/P
bulge once one forms is that we can use the actions to assign stellar
populations to the bulge.  Compared with pure $N$-body simulations,
simulations that include the physics of gas and star formation remain
expensive and do not permit rapid explorations of a large parameter
space, which are needed to test ideas of bulge formation.  With the
European Space Agency's {\it Gaia} mission now in its second data
release and the third data release imminent, such models are vital for
a detailed understanding of the formation of the MW's bulge.
Therefore a prescription for assigning stellar population properties
(ages, metallicities, abundances, \etc) to the star particles in pure
$N$-body simulations would be very valuable.  This action-based
metallicity prescription is different from past prescriptions based
solely on position and membership in the thin or thick disc
\citep[e.g.][]{bekki_tsujimoto11, martinez-valpuesta_gerhard13,
pdimatteo16} because stars at any particular location can be a mix of
very different metallicities \citep{hayden+15}.

This paper is organised as follows.  Section \ref{s:simulation}
presents the simulation suite used in this paper.  Section
\ref{s:actions} presents the evolution of stellar populations tagged
by actions in a fiducial simulation while Section \ref{s:simsuite}
generalises these results to the entire simulation suite.  In Section
\ref{s:chemistry} we consider prescriptions for assigning metallicity
to stars based on their initial actions.  We discuss our results and
present our conclusions in Section \ref{s:discussion}.  The
deconstruction of barred models by actions has never been presented
before.  Therefore in the online version of the paper the appendices
present figures for a standard analysis of all the other simulations.


\section{Simulations}
\label{s:simulation}

We use a suite of pure $N$-body simulations to track stars based on
their actions.  The models were set up using {\sc galactics}
\citep{widrow+08, widrow_dubinski05, kuijken_dubinski95}; they are
comprised of one or two discs in a truncated Navarro-Frenk-White (NFW)
dark matter halo \citep{nfw96}:
\begin{equation}
\rho(r) = \frac{2^{2-\gamma} \sigma_h^2}{4\pi a_h^2}
\frac{C(r)}{(r/a_h)^\gamma (1+r/a_h)^{3-\gamma}},
\end{equation}
\citep{widrow+08}, where the cutoff function $C(r)$ smoothly 
truncates the model at a finite radius:
\begin{equation}
C(r) =
\frac{1}{2}\mathrm{erfc}\left(\frac{r-r_h}{\sqrt{2}\delta r_h}\right).
\end{equation}
For all models we set $\sigma_h = 400\,\kms$, $a_h = 16.7\,\kpc$,
$\gamma = 0.873$, $r_h = 100\,\kpc$, and $\delta r_h = 25\,\kpc$.

The discs are exponential with an isothermal vertical profile:
\begin{equation}
\Sigma(R,z) = \Sigma_{0} \exp(-R/\Rd)~\mathrm{sech}^2 (z/\zd),
\end{equation}
where \Rd\ and \zd\ are the disc's scale-length and scale-height,
respectively.  For models 1-5 we set $\md = 2 \pi \Sigma_0 \Rd^2 =
5.2\times10^{10} \Msun$, and $\Rd = 2.4\,\kpc$.  Figure 1 of
\citet{debattista+17} shows the rotation curve of these systems, which
reveals that the galaxy is baryon-dominated out to $\sim 10\,\kpc$.
In this way the models resemble the MW \citep{cole_binney17}.  The
kinematics of the discs are set such that the radial velocity
dispersion, $\sigma_R$, decreases exponentially
\begin{equation}
\sigma_R^2(R) = \sigma_{R0}^2 \exp(-R/R_\sigma).
\end{equation}
We set $R_\sigma = 2.5\,\kpc$.  In models 1-5 we vary the radial and
vertical dispersions as listed in Table \ref{t:models}.  Our fiducial
model is model 2, while models 1 and 3 bracket it in radial random
motions, and 4 and 5 bracket it in vertical random motions (\ie\
thickness).  With these choices the minimum of the Toomre-$Q$ profiles
ranges from 0.8 to 1.5.
We use $6\times10^6$ particles in the disc and $4\times10^6$ particles
in the halo.  Thus disc particles have a mass $\simeq 1.1 \times 10^4
\Msun$, while halo particles have mass $\simeq 1.7\times 10^5 \Msun$.

\begin{table}
\begin{centering}
\begin{tabular}{ccccccc}\hline
\multicolumn{1}{c}{Model} &
\multicolumn{1}{c}{$\sigma_{R0}$} &
\multicolumn{1}{c}{$\zd$} & 
\multicolumn{1}{c}{$r_{\phi,R}$} &
\multicolumn{1}{c}{$r_{\phi,z}$} &
\multicolumn{1}{c}{$r_{R,z}$} \\ 
  & [\kms] & [\pc] & \\ \hline 
%
1 & 90 & 300 & -0.39 & -0.34 & 0.18 \\
2 & 128 & 300 & -0.41 & -0.35 & 0.19 \\
3 & 165 & 300 & -0.41 & -0.36 & 0.18 \\ 
4 & 128 & 150 & -0.40 & -0.37 & 0.20 \\
5 & 128 & 600 & -0.42 & -0.32 & 0.18 \\ 
T1 & $60+90$ & $100+400$ & -0.18 & -0.25 & 0.24 \\
T5 & $100+140$ & $300+900$ & -0.26 & -0.25 & 0.19 \\
HD1 & $45$ & $300$ & -0.38 & -0.32 & 0.16 \\
HD2 & $60$ & $300$ & -0.39 & -0.33 & 0.18 \\ \hline
\end{tabular}
\caption{The models used in this paper.  $\sigma_{R0}$ is the disc's 
central radial velocity dispersion(s), \zd\ is the disc scale
height(s), and $r_{i,j}$ is the Pearson $r$ for the correlation
between actions \act{i}\ and \act{j}.}
\label{t:models}
\end{centering}
\end{table}

Our simulation suite also includes two thin$+$thick disc models, T1
and T5, the former taken from \citet{debattista+17}.  The two discs
both have $\Rd = 2.4~\kpc$, each of mass $2.6\times10^{10}\Msun$, so
the total stellar mass is the same as in models 1-5.  In model T1, the
thin disc has $(\zd,\sig{R0},R_\sigma) = (100~\pc,60~\kms,4.5~\kpc)$
while for the thick disc these values are
$(400~\pc,90~\kms,2.5~\kpc)$.  The evolution of model T1 is described
in Section 3.2 of \citet{debattista+17}. In model T5, which has not
been presented elsewhere, we set a thin disc with
$(\zd,\sig{R0},R_\sigma) = (300~\pc,100~\kms,4.5~\kpc)$ while for the
thick disc these parameters are $(900~\pc,140~\kms,3.5~\kpc)$.  The
dark matter halo of both models is the same as that in models 1-5.

\begin{figure}
\centerline{
\includegraphics[angle=0.,width=\hsize]{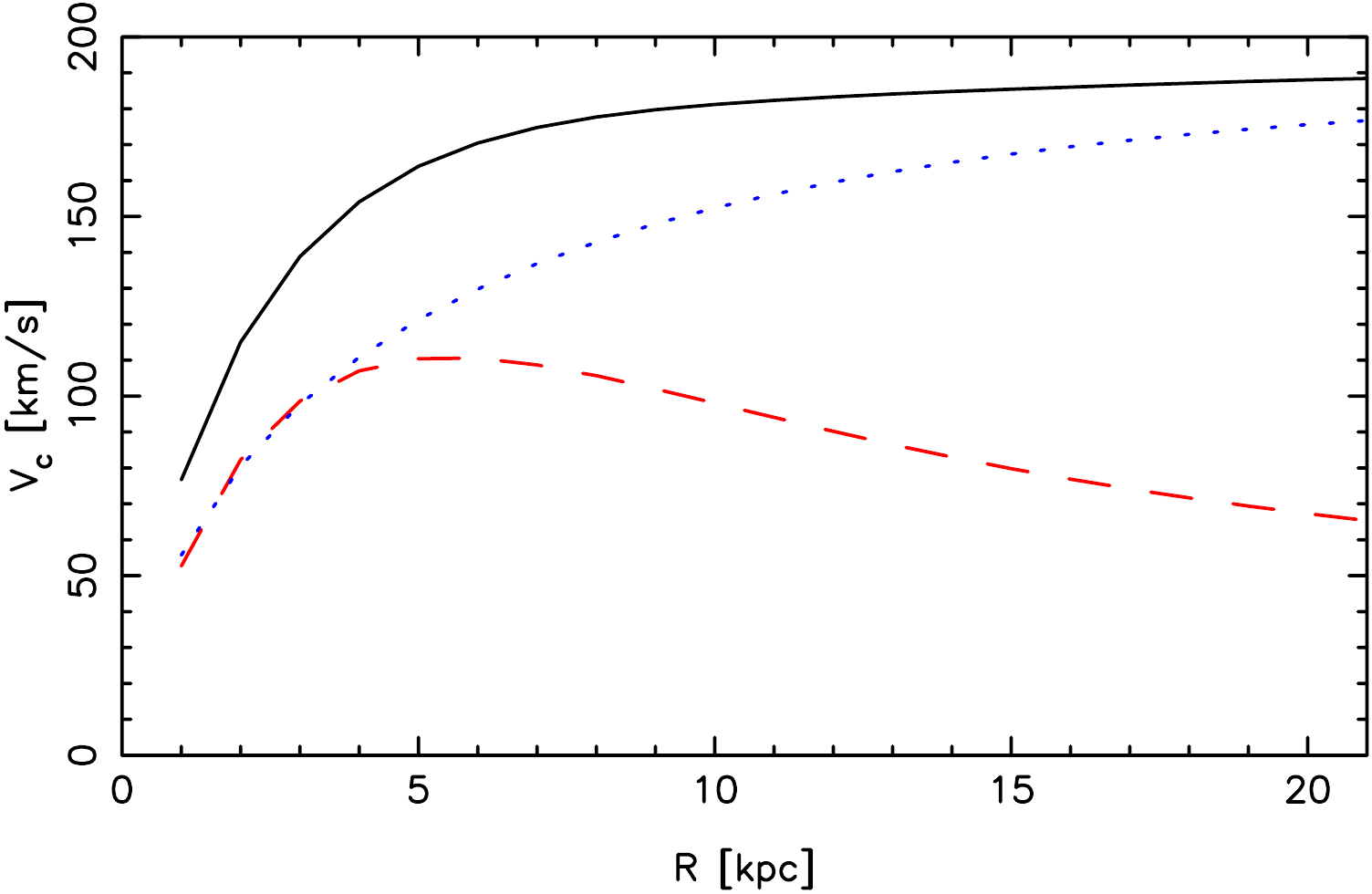}
}
\caption{The rotation curve of the halo-dominated models, HD1 and HD2.  
The solid (black) line shows the total rotation curve while the dashed
(red) and dotted (blue) lines show the disc and halo contributions,
respectively.
\label{f:rotcurvHD}}
\end{figure}

We also include two dark matter-dominated systems, models HD1 and HD2
(``halo dominated''), in which the dark matter and baryons contribute
about equally to the inner rotation curve, as shown in
Fig. \ref{f:rotcurvHD}.  In these models we keep the same halo density
profile as in the other models, and the same disc scale-length, \Rd,
scale-height, \zd, and dispersion scale-length, $R_\sigma$, as in the
fiducial model.  The only two differences are that we lower the mass
of the disc to $35\%$ of the fiducial model and lower the central
velocity dispersion to have Toomre-$Q$ profiles similar to model 1
(HD1) and model 2 (HD2).

The models were evolved with {\sc pkdgrav} \citep{pkdgrav}, using a
particle softening of $\epsilon = 50~\pc$ and $\epsilon = 100~\pc$ for
star and halo particles, respectively, in all the simulations.  The
base timestep is $\Delta t = 5\Myr$, and is refined such that each
particle's timestep satisfies $\delta t = \Delta t/2^n < \eta
\sqrt{\epsilon/a_g}$, where $a_g$ is the acceleration at the
particle's current position. We use $\eta = 0.2$ and set the opening
angle of the tree code gravity calculation $\theta = 0.7$.  We evolve
the models for $5\Gyr$, except for models HD1 and HD2, which we
evolve for $10\Gyr$ since their bars take longer to form.  During this
time a bar forms and buckles, resulting in a B/P bulge.  At the end of
the simulations we align the bars along the $x$-axis to facilitate our
analysis and inter-comparison.

\begin{figure*}
\centerline{
\includegraphics[angle=0.,width=0.3\hsize]{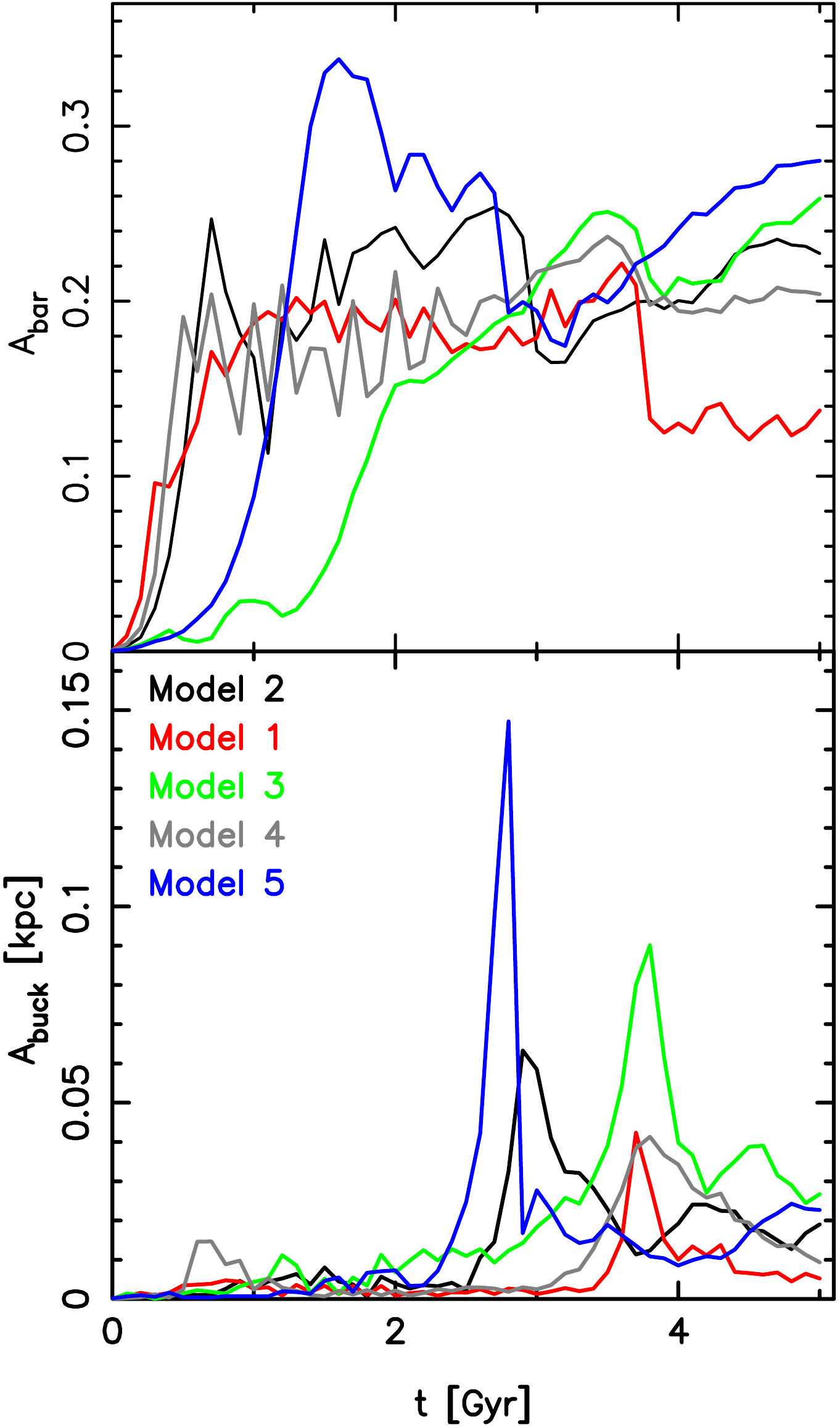}
\includegraphics[angle=0.,width=0.3\hsize]{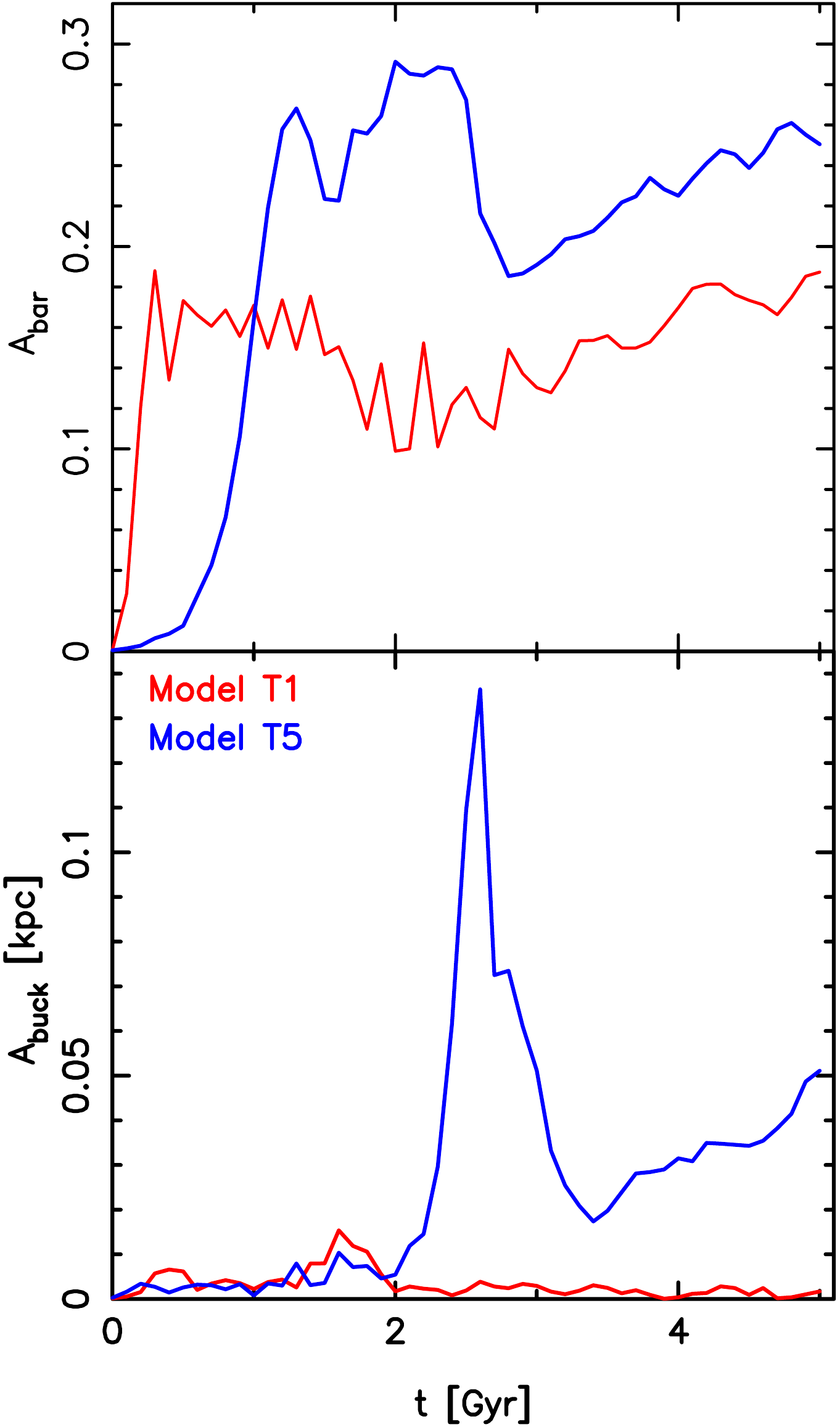}
\includegraphics[angle=0.,width=0.3\hsize]{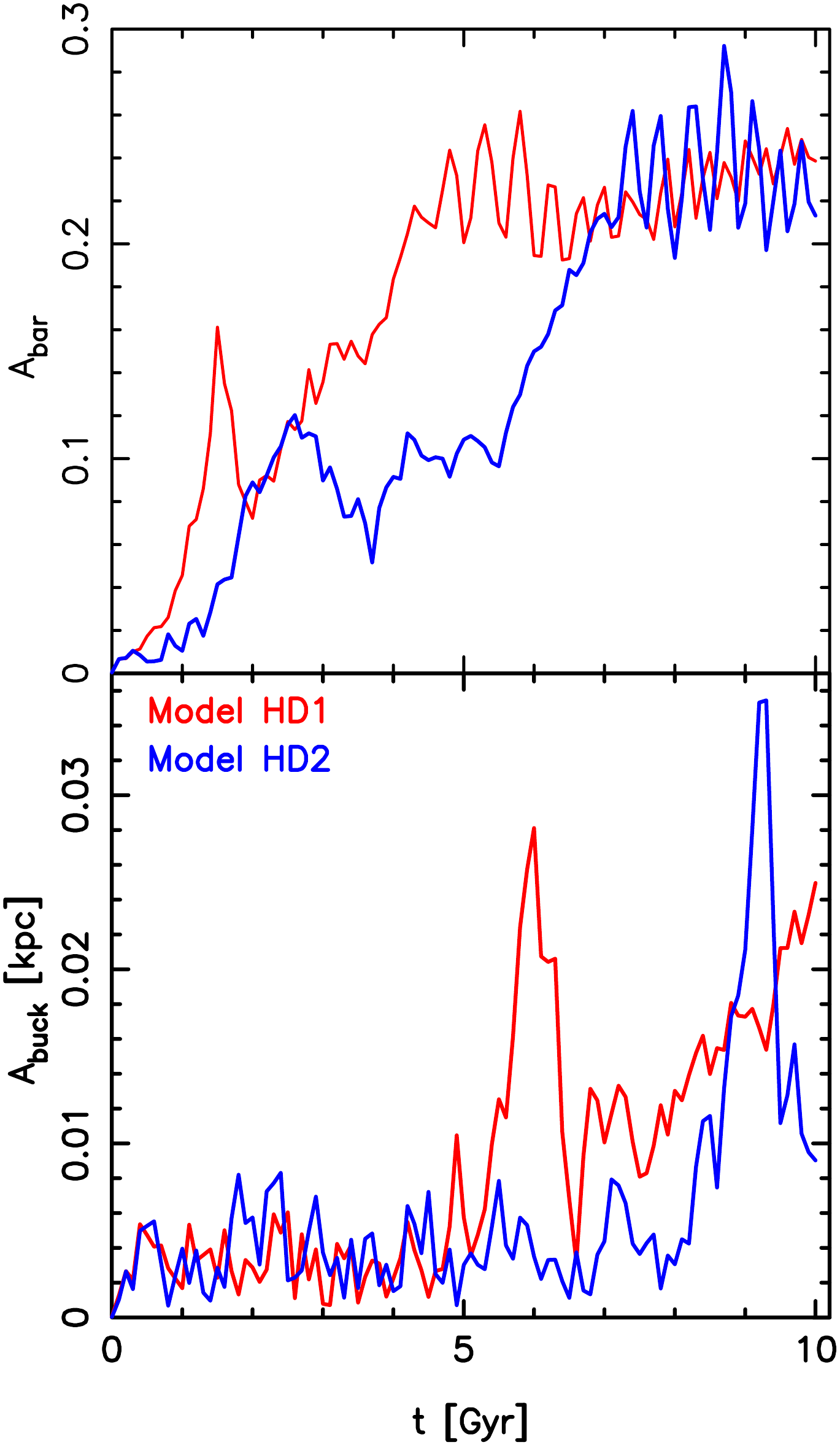}\
}
\caption{The global bar (top) and buckling (bottom) amplitudes of the 
simulation suite.  Note that both vertical and horizontal axes change
between the different panels.  All models form a bar and buckle, but
there is a wide range of buckling strengths.
\label{f:suite}}
\end{figure*}

Fig. \ref{f:suite} presents the evolution of the bar and the buckling
global amplitudes of the full simulation suite.  We define these
amplitudes as:
\begin{equation}
A_\mathrm{bar} = \left| \frac{\sum_{k} m_k e^{2 i \phi_k}}{\sum_{k} m_k} \right|,
\end{equation}
and
\begin{equation}
A_\mathrm{buck} = \left| \frac{\sum_{k} z_k m_k e^{2 i \phi_k}}{\sum_{k} m_k} \right|,
\end{equation}
where $m_k$, $z_k$ and $\phi_k$ are the mass, vertical position and
azimuthal angle of particle $k$ and the sums are over all star
particles \citep[e.g.][]{sellwood_athanassoula86, debattista+06}.
The models exhibit a range of bar growth rates, and bar strengths.
All bars weaken at buckling. By the end of the simulations most of the
bars have started to grow again.  The fiducial model 2 forms a bar at
$t\sim 1\Gyr$, and buckles at $t\sim 3\Gyr$.  The cooler model 1 forms
its bar faster and buckles at $t\sim 3.7\Gyr$.  The hotter model 3
forms its bar later still, but still buckles at $t\sim 3.7\Gyr$.  The
thinner (thicker) model 4 (5) forms earlier (later).  Buckling in
model 5 is the strongest of the entire simulation suite.  The bar in
model T1 forms quite rapidly and buckles quite mildly, whereas model
T5 forms a bar somewhat later and undergoes a strong buckling.  Model
HD1 forms a bar at $\sim 4\Gyr$, and buckles at $6\Gyr$.  After $t\sim
7\Gyr$ $A_\mathrm{buck}$ rises monotonically but this reflects a bend
in the bar rather than a buckling.  Finally model HD2 forms a bar at
$7\Gyr$ and buckles at $t\sim 9.2\Gyr$.

\begin{figure*}
\centerline{
\includegraphics[angle=0.,width=0.5\hsize]{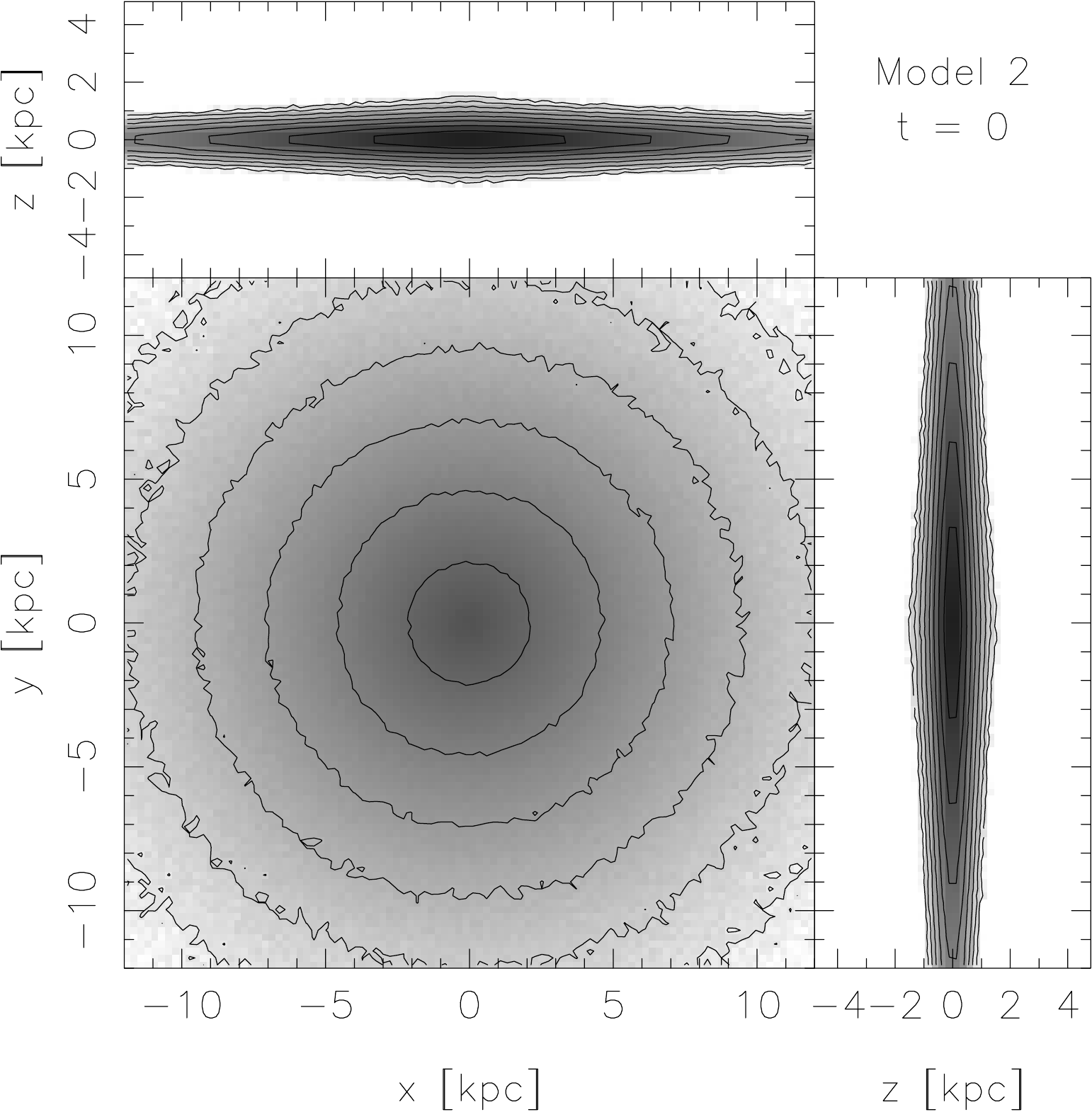}
\includegraphics[angle=0.,width=0.5\hsize]{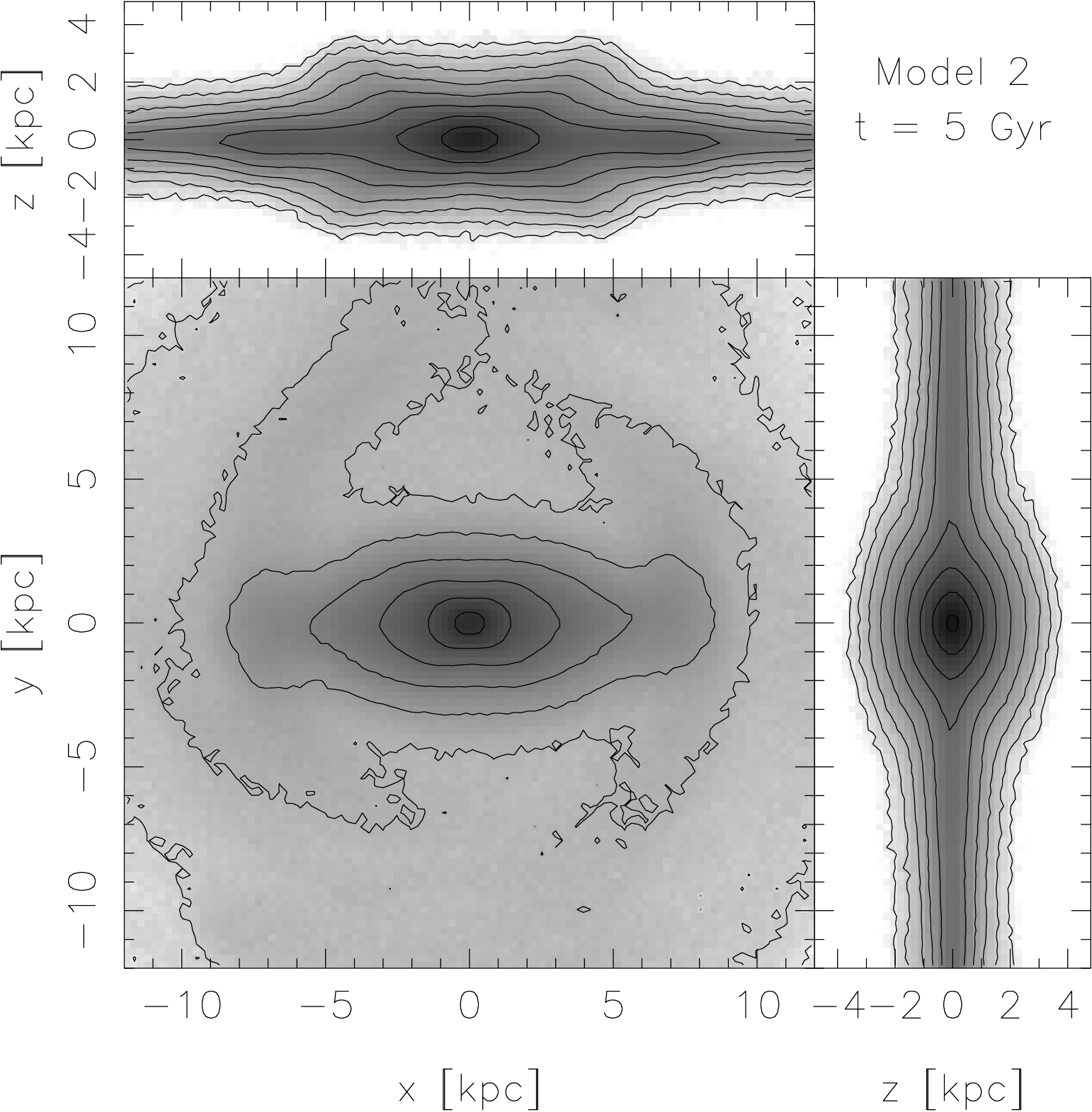}\
}
\caption{The density distribution of model 2 at $t=0$ (left) 
and at the end of the simulation, at $t=5\Gyr$, (right) for 3
different projections.  The density levels are arbitrary.  A prominent
bar, with a B/P-shaped bulge, has formed by $t=5\Gyr$.  The bar has
been rotated into the $x$-axis.
\label{f:model}}
\end{figure*}

Fig. \ref{f:model} presents a view of model 2 at the start and end of
the simulation.  The main analysis presented will focus on this model.
Plots similar to those for model 2 for the rest of the simulation
suite are presented in the Appendices of the online edition.


\section{Evolution in initial action space}
\label{s:actions}

We use {\sc agama} \citep{agama} to measure the actions in the radial,
vertical and azimuthal directions at $t=0$. We compute the potential
of the disc and halo separately, assuming spherical symmetry for the
halo and a flattened axisymmetric distribution for the disc.  We have
verified that the total potential computed this way reproduces the
rotation curve.  We then compute the actions assuming the local
axisymmetric St\"ackel fudge \citep{binney12}.

We treat the actions computed at $t=0$ as tags for stellar particles
at the end of the simulation.  We stress that at the end of the
simulations \emph{these are still the actions computed for the initial
conditions}.  In essence we are tracing where stars end up based on
their initial actions.  In order to make this evident at all times, we
subscript actions with a `0'.  Thus our notation for radial, azimuthal
and vertical action is $(\act{R},\act{\phi},\act{z})$.  This
\emph{does not} mean that we are assuming that the actions are
conserved during the process of bar and B/P bulge formation.

Since, for a finite disc, the maximum values of \act{R}\ and
\act{z}\ depend on the radius within the galaxy, and thus on
\act{\phi}, the actions may be correlated to some extent. Table 
\ref{t:models} lists the Pearson $r$ values of the correlations 
between all pairs of actions in the various simulations.  The largest
absolute Pearson $r$ value is 0.42 for the correlation between
\act{\phi}\ and \act{R} in model 5.  Frequencies, on the other hand,
are much more covariant, with the Pearson $r$ between the radial
frequency, \omr, and the vertical frequency, \omz, of 0.55 in the
fiducial model, and 0.70 in model 5.  For this reason we do not
consider the frequencies in any of our analysis.

Part of our analysis consists of splitting the models into quartiles
of their action distributions.  The ranges of actions in these
quartiles for each model are listed in Table~\ref{t:quartiles}.

\begin{table*}
\begin{centering}
\begin{tabular}{cccccc}\hline
\multicolumn{1}{c}{Model} &
\multicolumn{1}{c}{Action} &
\multicolumn{1}{c}{1$^{st}$ quartile} &
\multicolumn{1}{c}{2$^{nd}$ quartile} &
\multicolumn{1}{c}{3$^{rd}$ quartile} &
\multicolumn{1}{c}{4$^{th}$ quartile} \\
 & & [\kmskpc] & [\kmskpc] & [\kmskpc] & [\kmskpc] \\ \hline 
1  & \act{\phi} & $<348$ & $348$-$822$ & $822$-$1473$ & $>1473$ \\
2  & \act{\phi} & $<255$ & $255$-$731$ & $731$-$1415$ & $>1415$ \\
3  & \act{\phi} & $<163$ & $163$-$568$ & $568$-$1315$ & $>1315$ \\
4  & \act{\phi} & $<265$ & $265$-$736$ & $736$-$1406$ & $>1406$ \\
5  & \act{\phi} & $<234$ & $234$-$712$ & $712$-$1414$ & $>1414$ \\ 
T1 & \act{\phi} & $<336$ & $336$-$768$ & $768$-$1394$ & $>1394$ \\ 
T5 & \act{\phi} & $<223$ & $223$-$639$ & $639$-$1282$ & $>1282$ \\
HD1& \act{\phi} & $<278$ & $278$-$626$ & $626$-$1139$ & $>1139$ \\
HD2& \act{\phi} & $<258$ & $258$-$609$ & $609$-$1127$ & $>1127$ \\ \hline
1  &    \act{R} & $<4$ & $4$-$12$ & $12$-$27$ & $>27$ \\
2  &    \act{R} & $<9$ & $9$-$27$ & $27$-$65$ & $>65$ \\ 
3  &    \act{R} & $<17$ & $17$-$53$ & $53$-$130$ & $>130$ \\
4  &    \act{R} & $<9$ & $9$-$27$ & $27$-$64$ & $>64$ \\ 
5  &    \act{R} & $<9$ & $9$-$27$ & $27$-$67$ & $>67$ \\
T1 &    \act{R} & $<7$ & $7$-$19$ & $19$-$44$ & $>44$ \\
T5 &    \act{R} & $<18$ & $18$-$48$ & $48$-$106$ & $>106$ \\
HD1&    \act{R} & $<1$ & $1$-$4$ & $4$-$9$ & $>9$ \\
HD2&    \act{R} & $<2$ & $2$-$7$ & $7$-$16$ & $>16$ \\ \hline
1  &    \act{z} & $<1.9$ & $1.9$-$5.0$ & $5.0$-$11.5$ & $>11.5$ \\
2  &    \act{z} & $<1.8$ & $1.8$-$5.0$ & $5.0$-$11.6$ & $>11.6$ \\
3  &    \act{z} & $<1.8$ & $1.8$-$4.9$ & $4.9$-$11.7$ & $>11.7$ \\
4  &    \act{z} & $<0.6$ & $0.6$-$1.8$ & $1.8$-$4.2$ & $>4.2$ \\ 
5  &    \act{z} & $<5.3$ & $5.3$-$13.9$ & $13.9$-$31.4$ & $>31.4$ \\ 
T1 &    \act{z} & $<0.7$ & $0.7$-$2.2$ & $2.2$-$7.9$ & $>7.9$ \\
T5 &    \act{z} & $<3.2$ & $3.2$-$10$ & $10$-$28$ & $>28$ \\
HD1&    \act{z} & $<1.2$ & $1.2$-$3.3$ & $3.3$-$7.4$ & $>7.4$ \\
HD2&    \act{z} & $<1.2$ & $1.2$-$3.3$ & $3.3$-$7.5$ & $>7.5$ \\ \hline  \hline
\end{tabular}
\caption{Action quartiles of the models.}
\label{t:quartiles}
\end{centering}
\end{table*}

\subsection{Dependence of the  bar strength on the actions}
\label{ss:barstrength}

\begin{figure*}
\centerline{
\includegraphics[angle=0.,width=0.25\hsize]{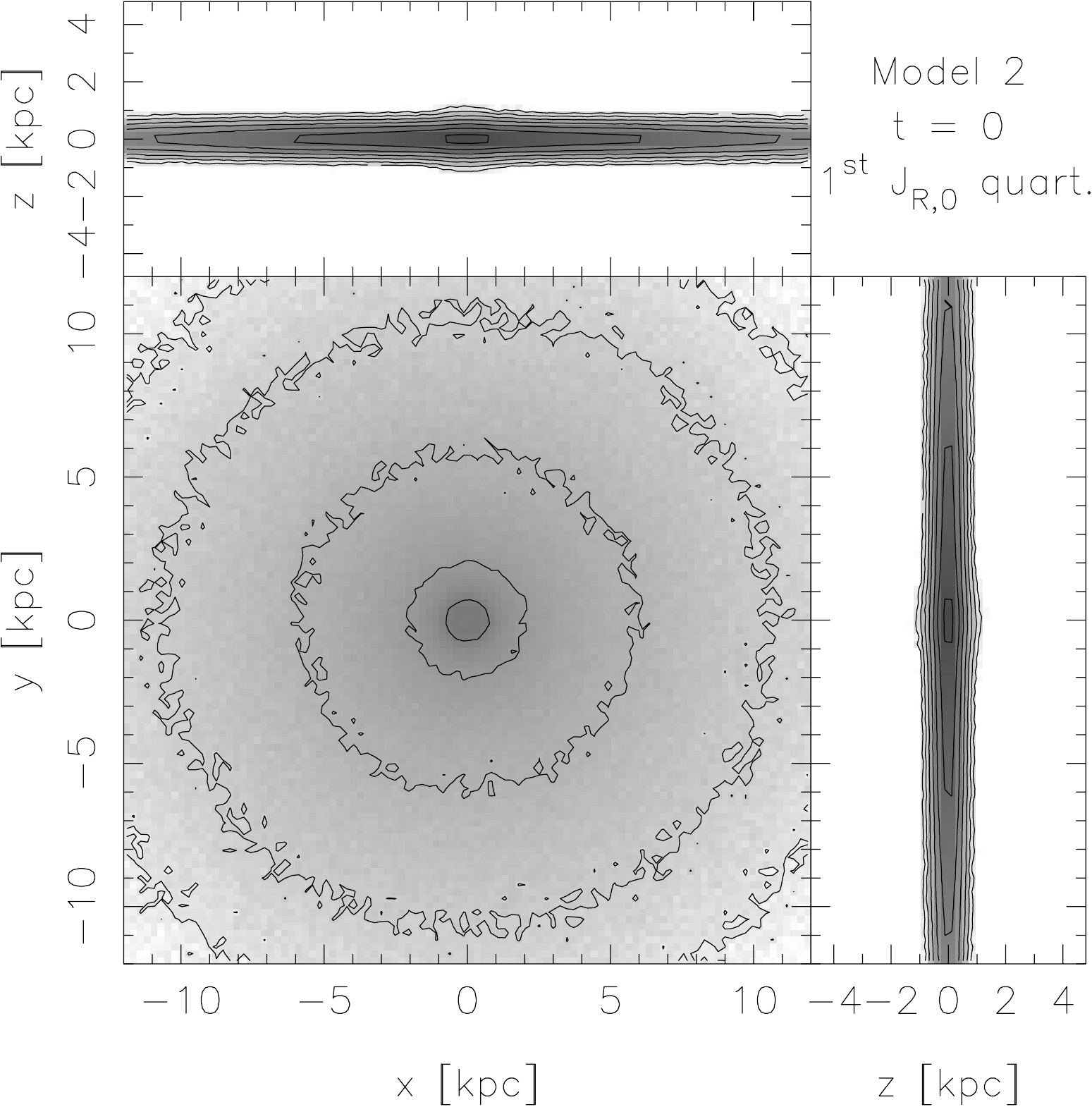}
\includegraphics[angle=0.,width=0.25\hsize]{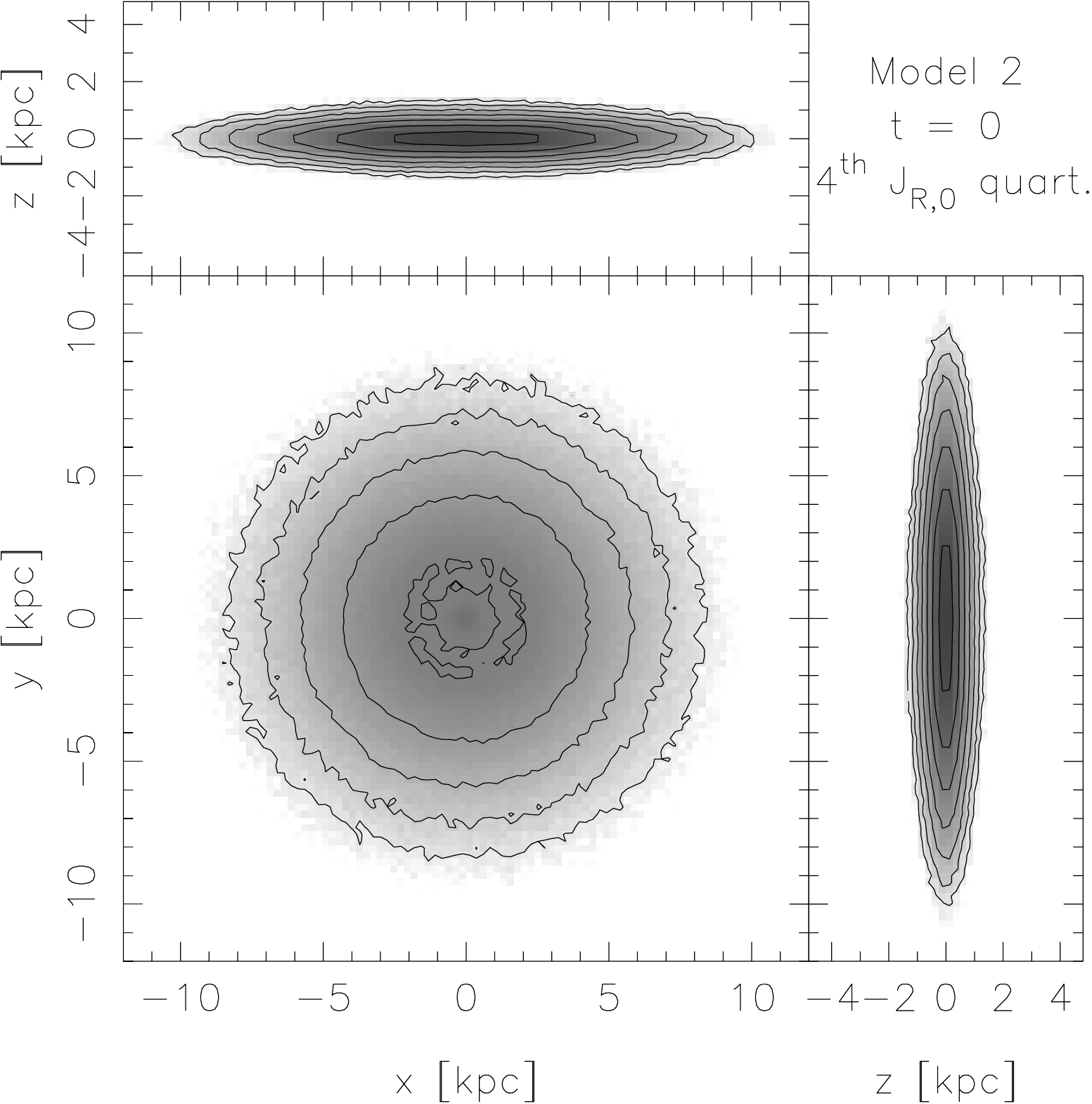}
\includegraphics[angle=0.,width=0.25\hsize]{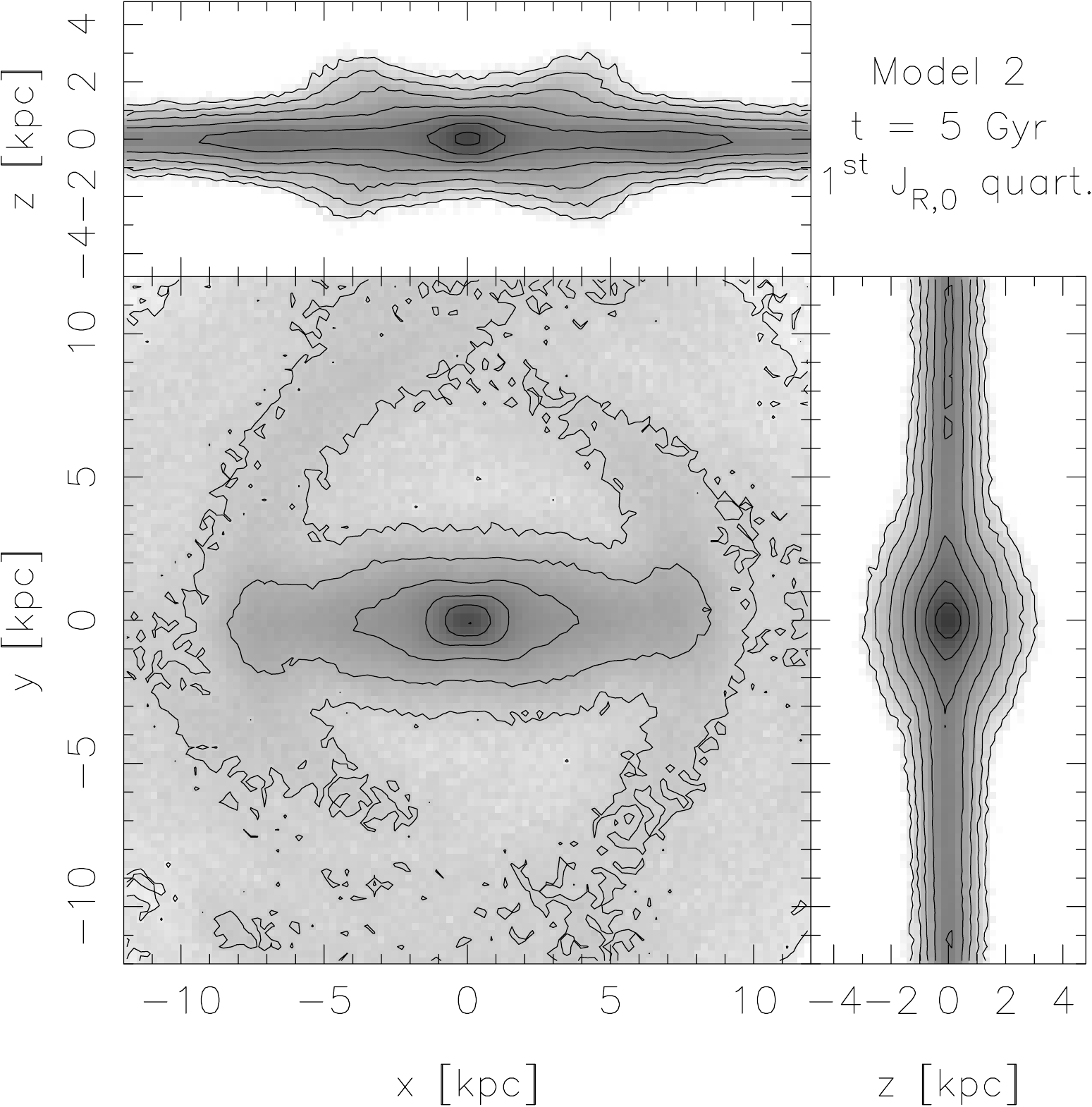}
\includegraphics[angle=0.,width=0.25\hsize]{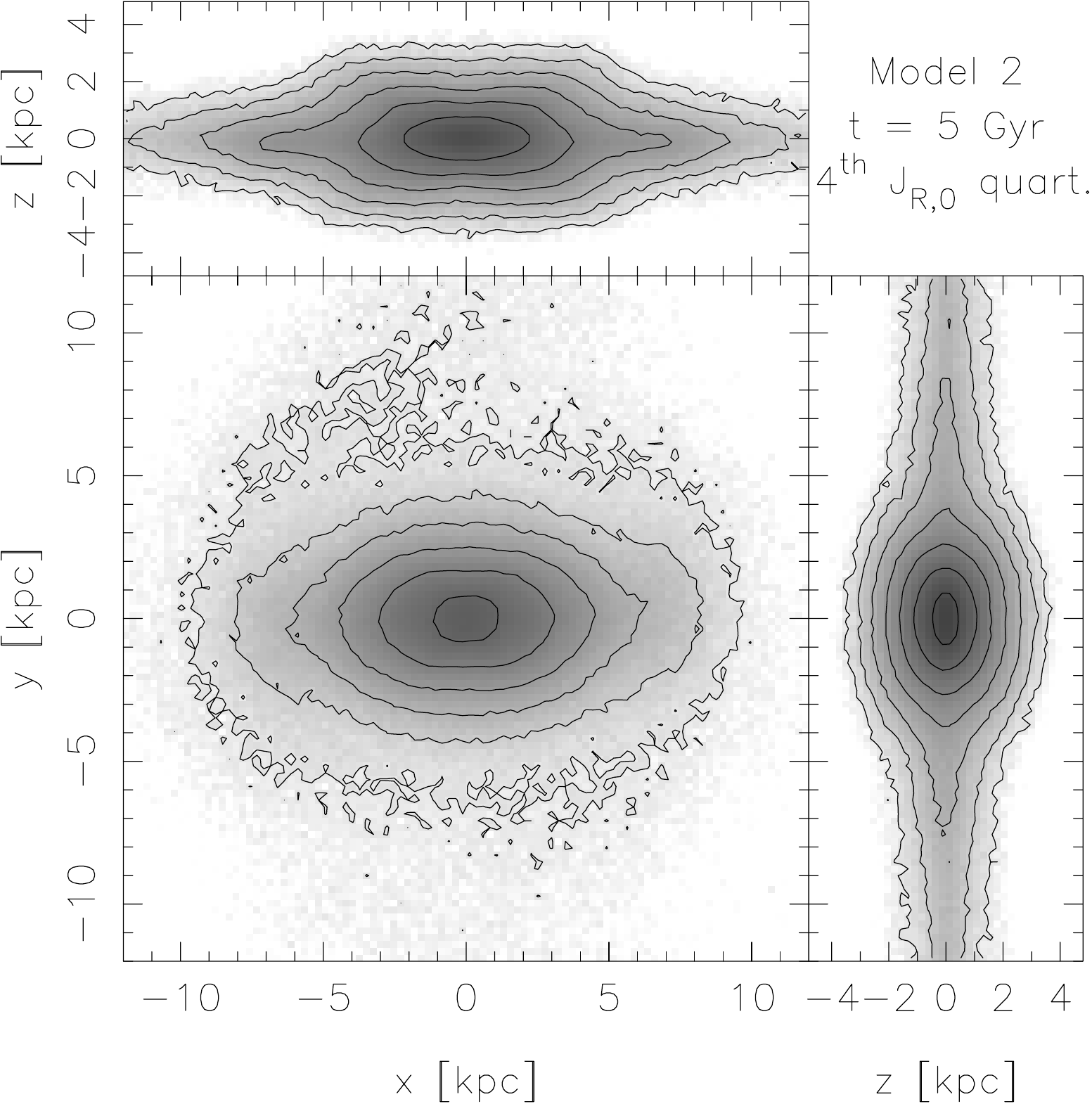}\
}
\centerline{
\includegraphics[angle=0.,width=0.25\hsize]{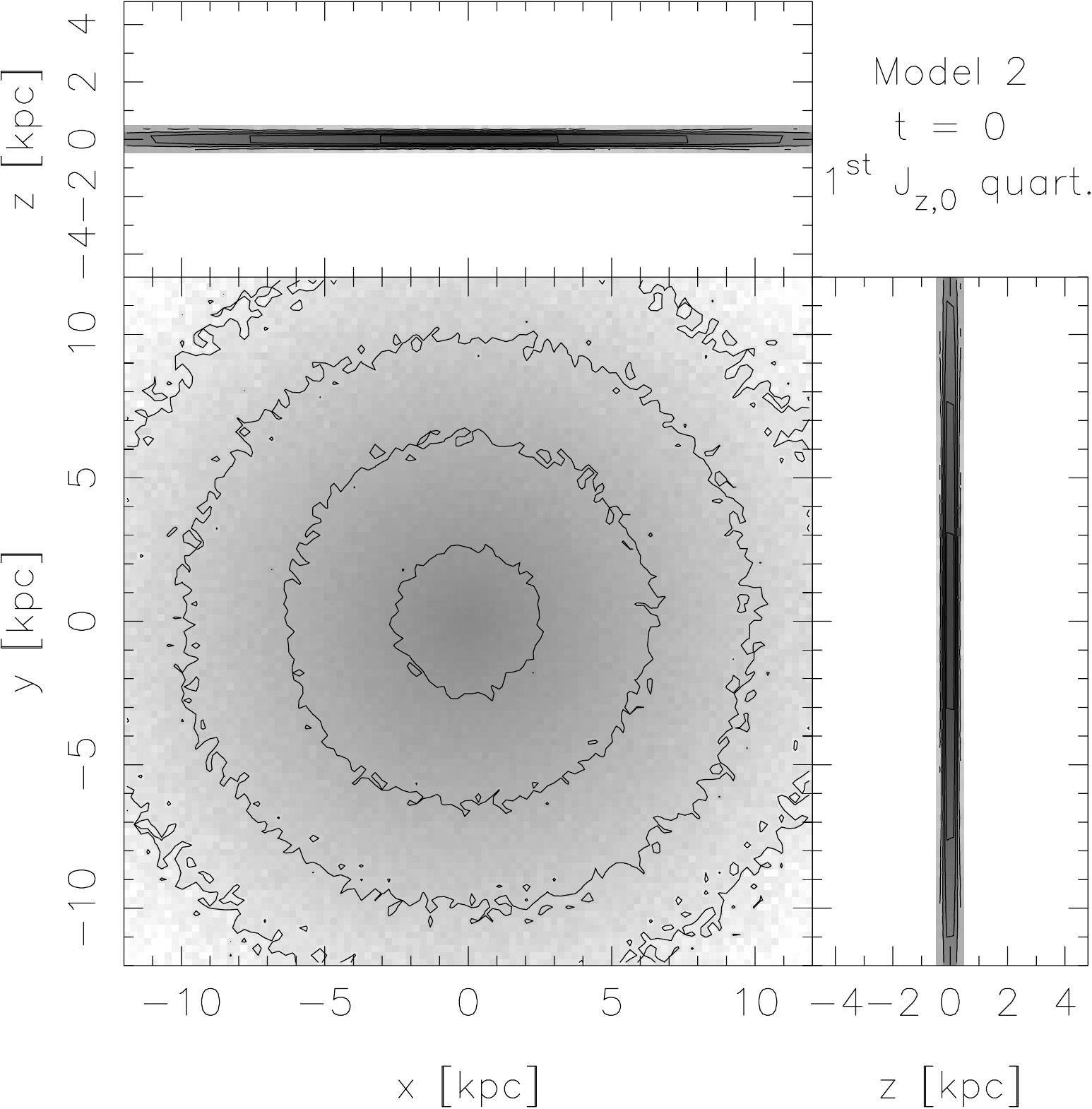}
\includegraphics[angle=0.,width=0.25\hsize]{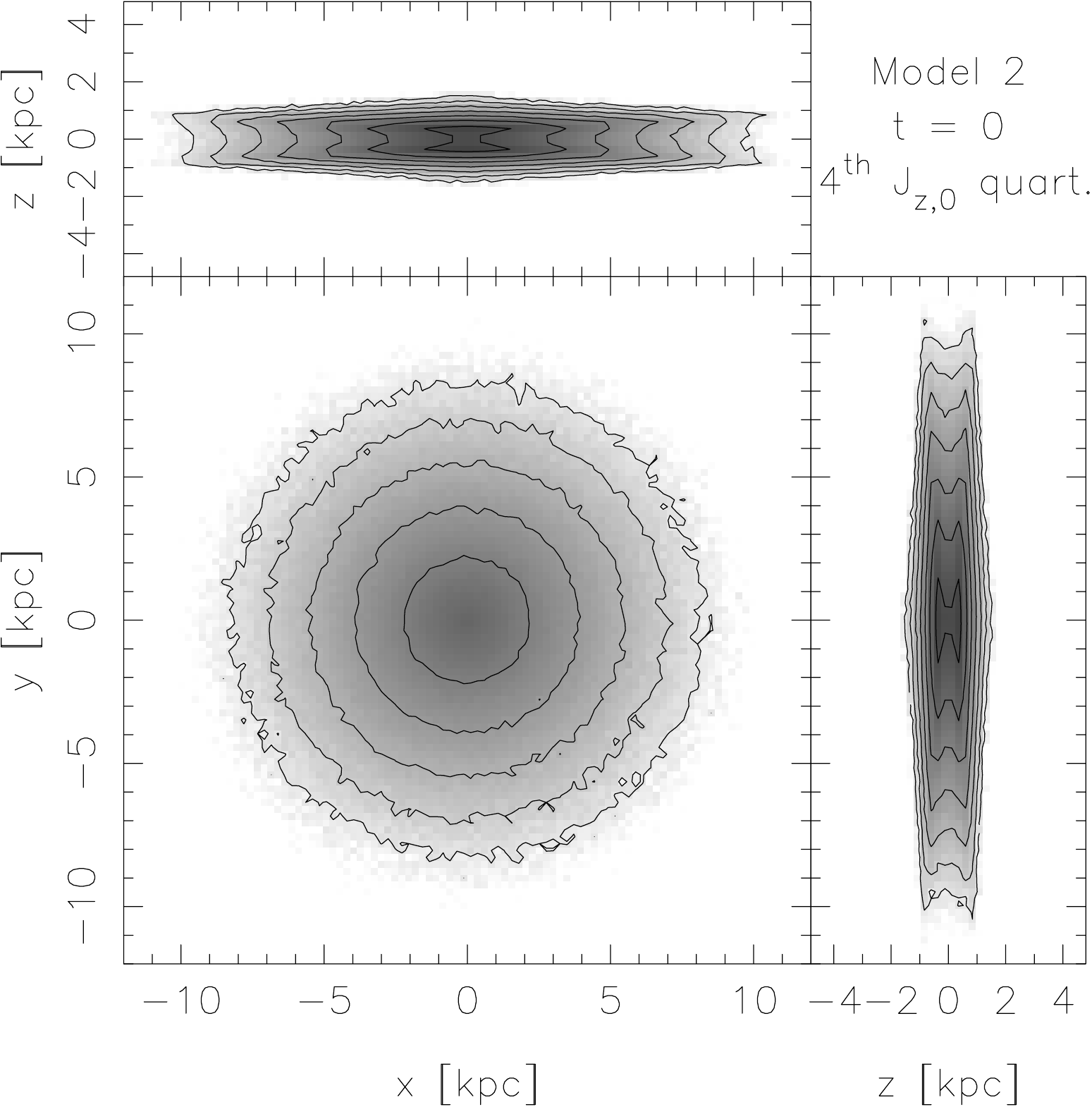}
\includegraphics[angle=0.,width=0.25\hsize]{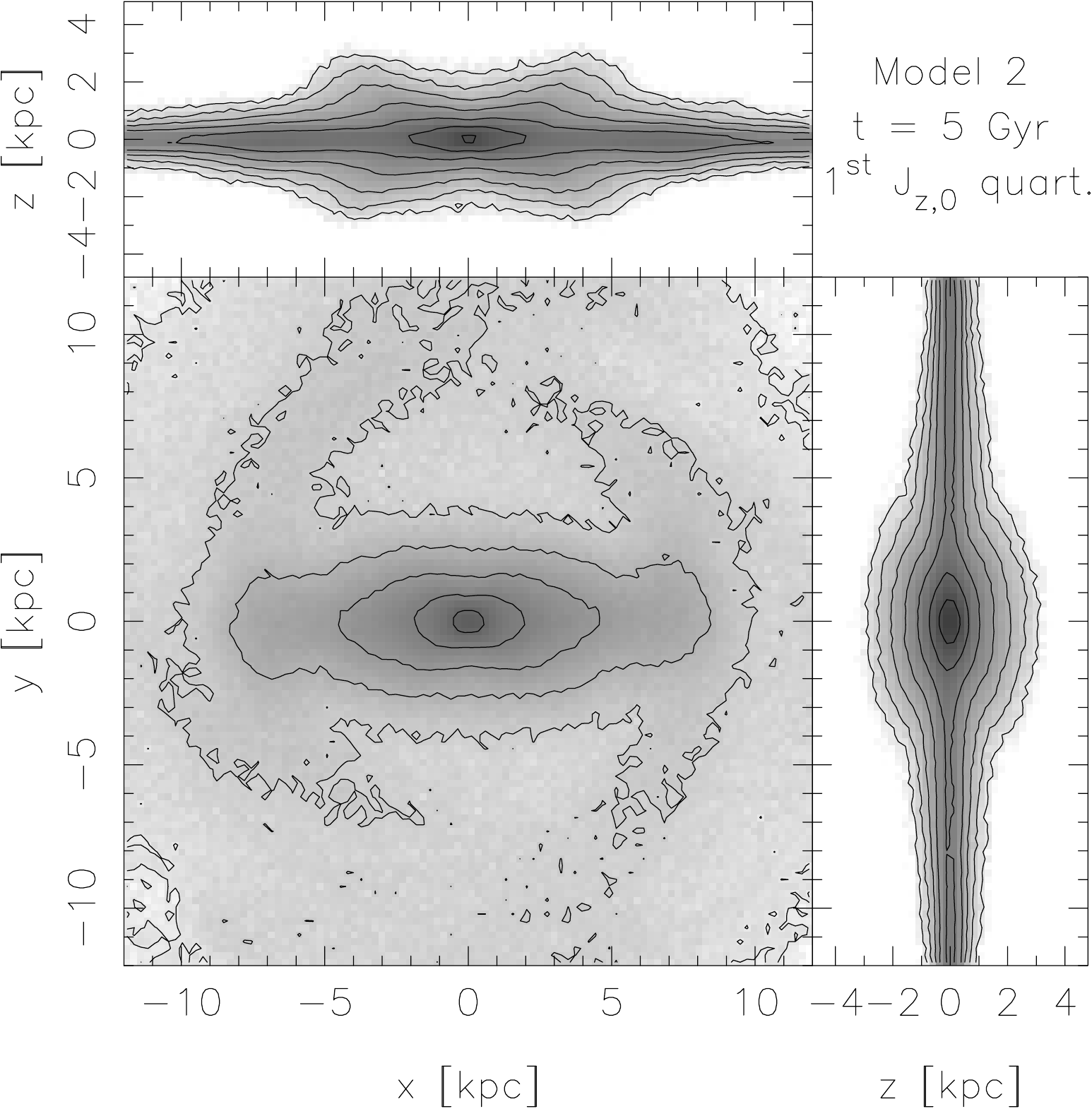}
\includegraphics[angle=0.,width=0.25\hsize]{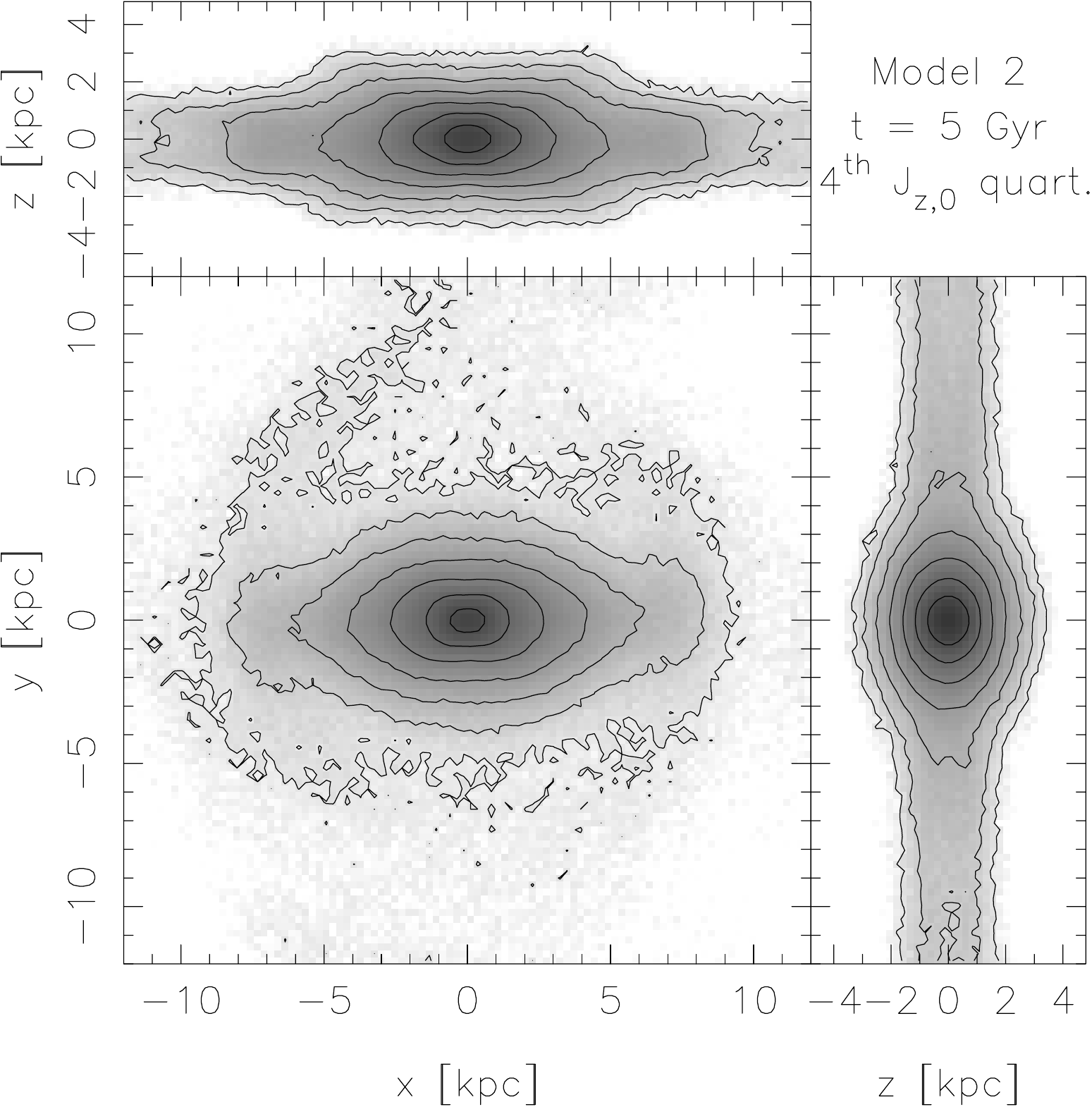}\
}
\centerline{
\includegraphics[angle=0.,width=0.25\hsize]{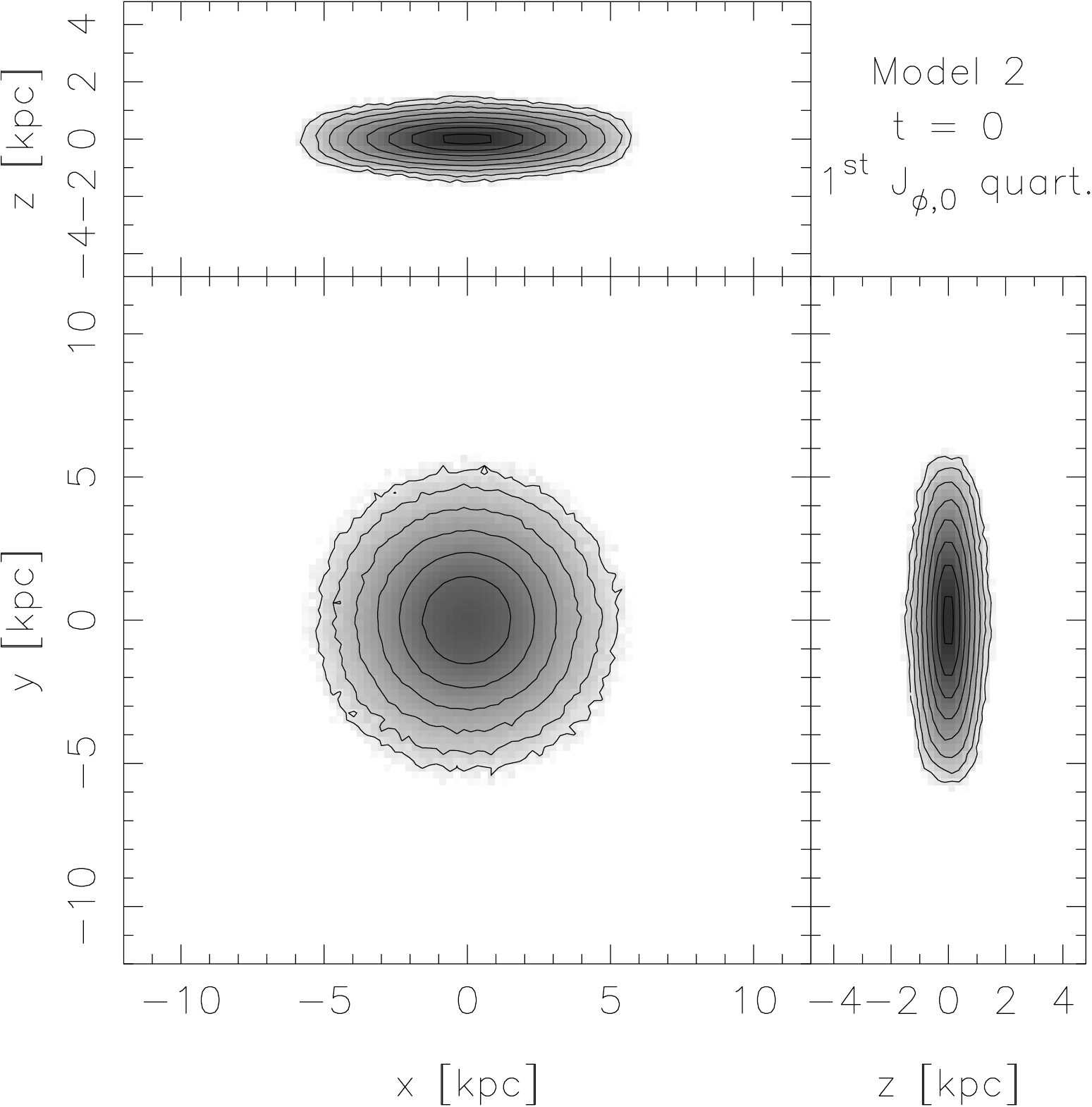}
\includegraphics[angle=0.,width=0.25\hsize]{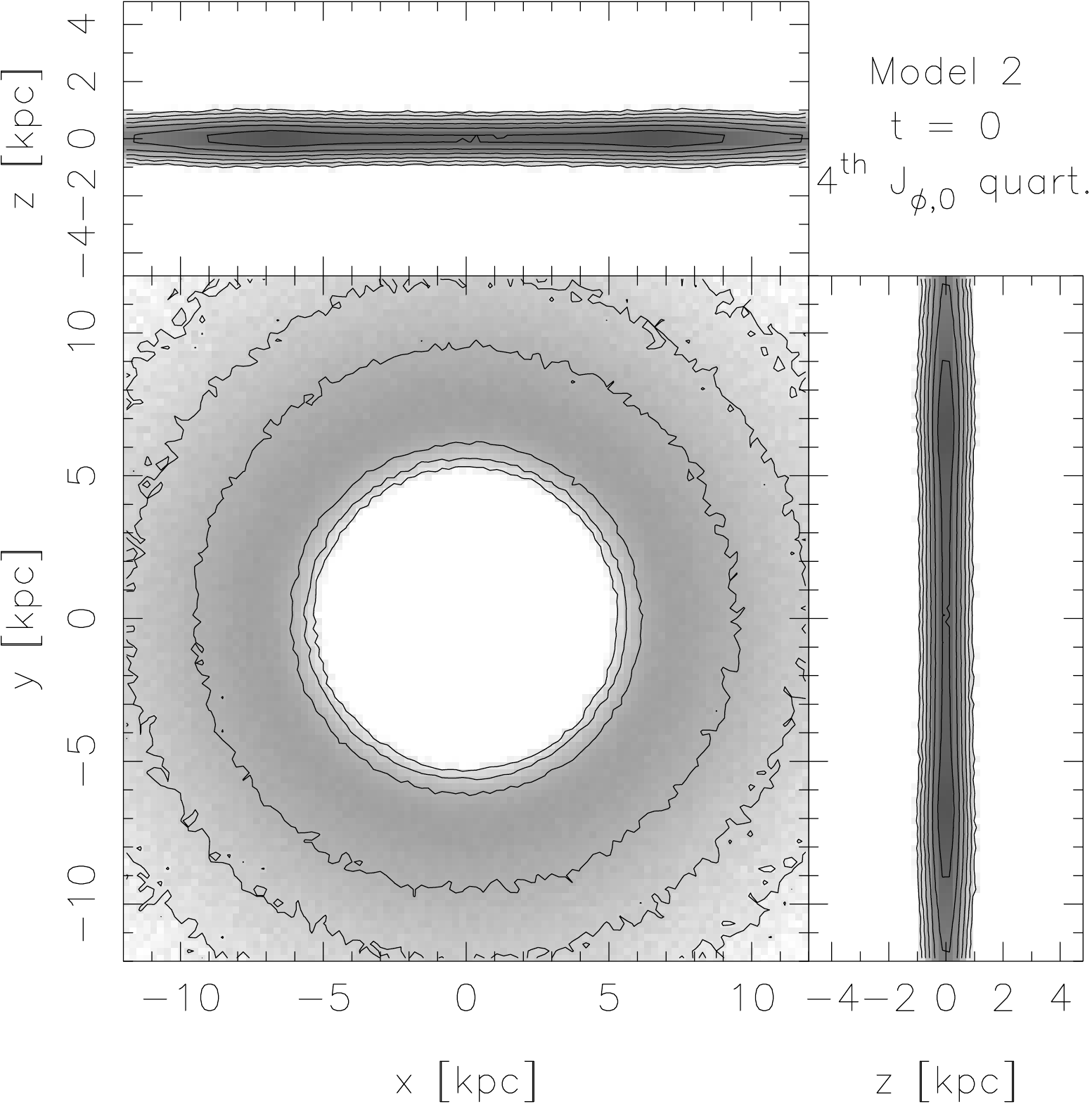}
\includegraphics[angle=0.,width=0.25\hsize]{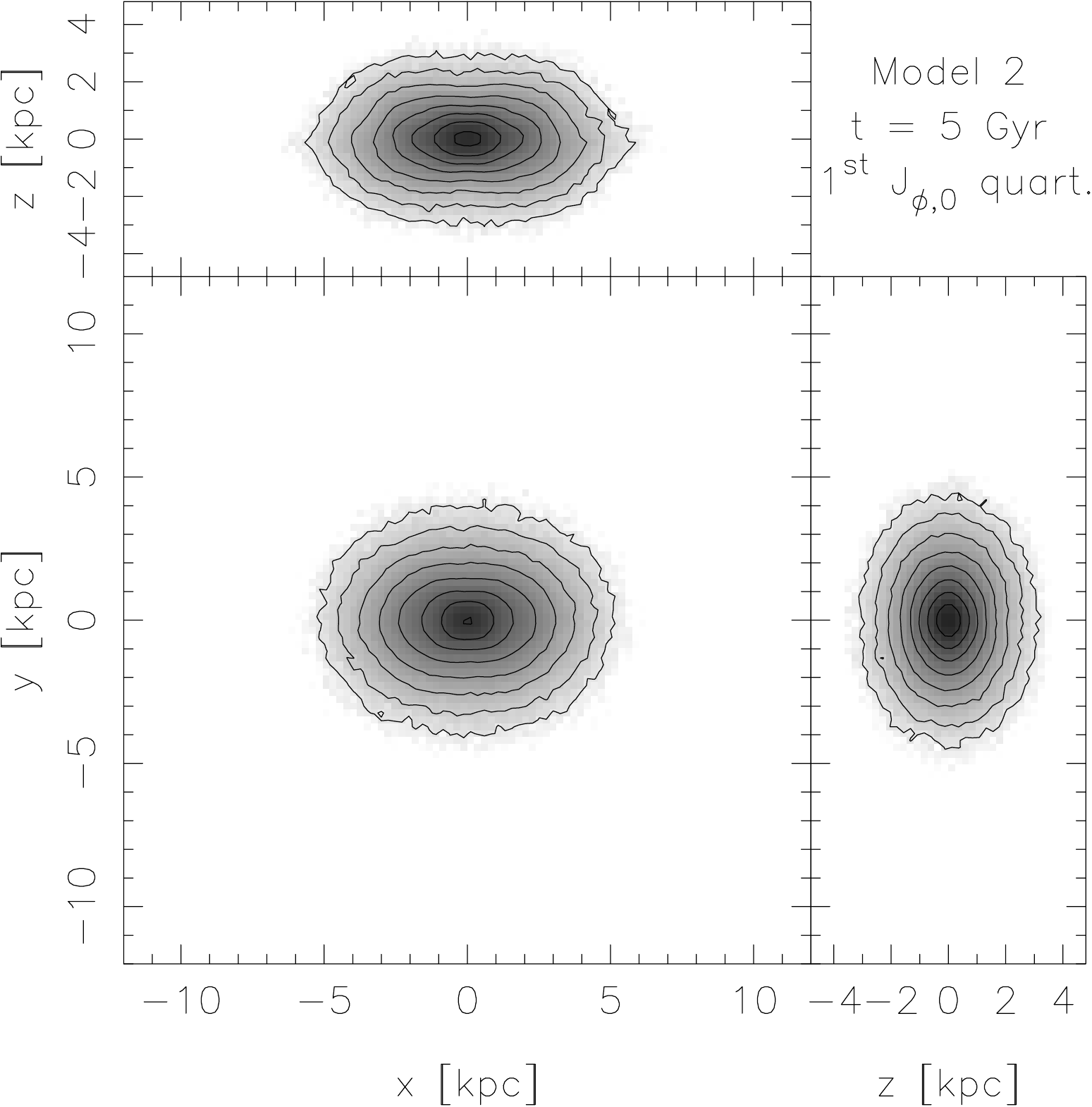}
\includegraphics[angle=0.,width=0.25\hsize]{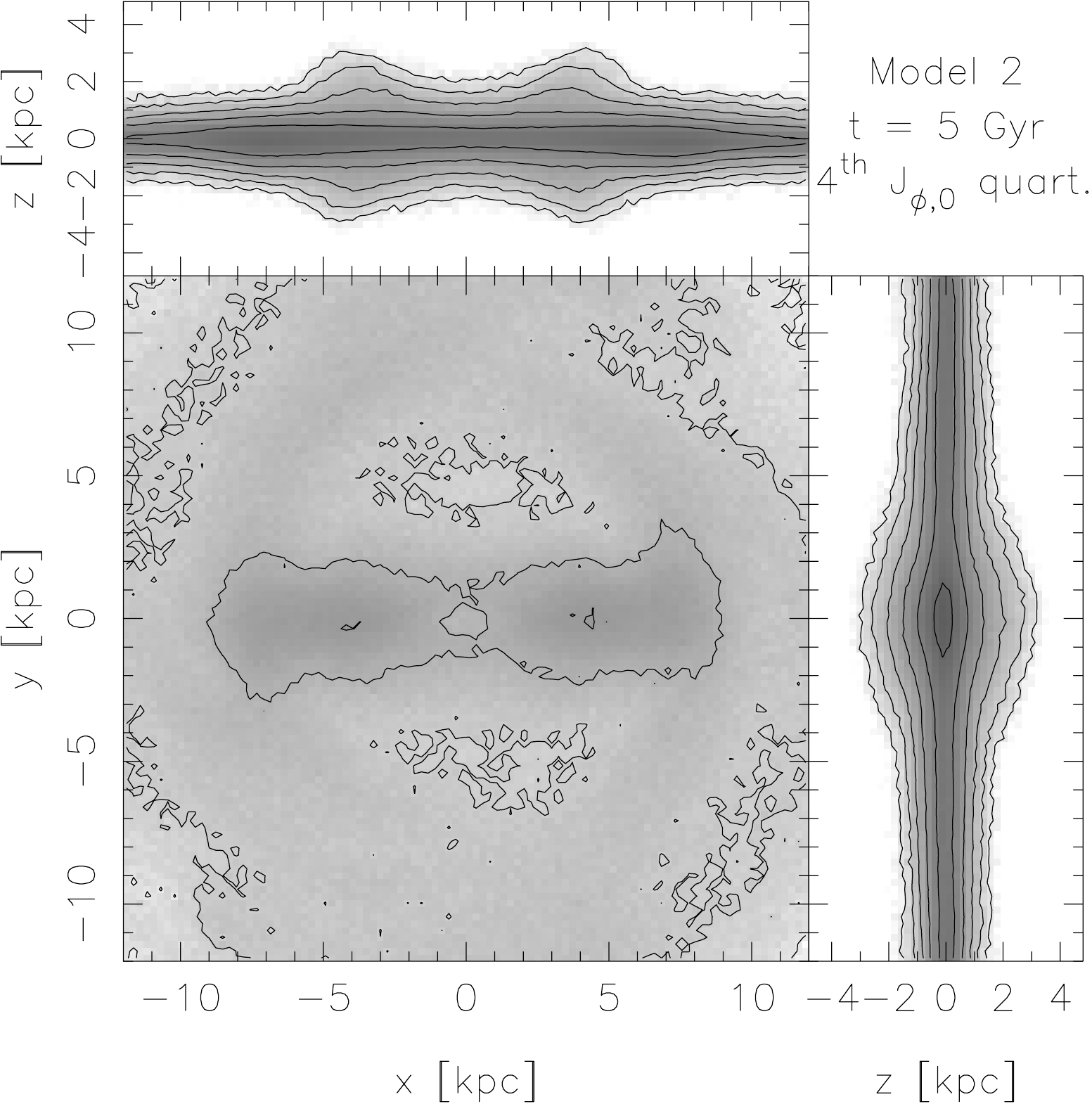}\
}
\caption{Evolution of the mass distribution in model 2 separated into 
action quartiles.  At top is \act{R}, followed by \act{z}, while at
bottom is \act{\phi}.  The first and third columns show the lowest
quartile while the second and fourth columns show the highest
quartile.  The two left columns show $t=0$, while the two right
columns show the end of the simulation, at $5\Gyr$.
\label{f:actionquartiles}}
\end{figure*}

Fig. \ref{f:actionquartiles} shows the initial and final density
distributions in the lowest and highest quartiles of each action
(listed in Table \ref{t:quartiles}). The top row shows the evolution
of the \act{R}\ quartiles.  Initially the population with the largest
\act{R}\ is concentrated towards the centre, while those in the lowest
\act{R}\ quartile are more radially extended.  This continues to be
true once the bar forms.  At $t=5\Gyr$ the bar is visibly stronger in
the lowest \act{R}\ quartile.  The middle row reveals a very similar
behaviour for \act{z}\ both before and after the bar forms.  Compared
with the bar in the \act{R}\ quartiles, the bar is weaker in the
lowest \act{z}\ quartile and stronger at the highest one, \ie\ there
is a larger range in the bar strength when populations are separated
by \act{R}\ than when they are separated by \act{z}.  The radial
action therefore seems to be a more important factor in determining
the bar strength of a given population than is the vertical action.
The third row shows \act{\phi}\ which reveals that \act{\phi}\
increases with radius, as expected.  The lowest \act{\phi}\ quartile
is almost completely part of the bar at $5\Gyr$ but is less
peanut-shaped viewed edge-on, while less of the highest \act{\phi}\
quartile ends up in the bar but is peanut shaped when viewed edge-on.

\begin{figure}
\centerline{
\includegraphics[angle=0.,width=\hsize]{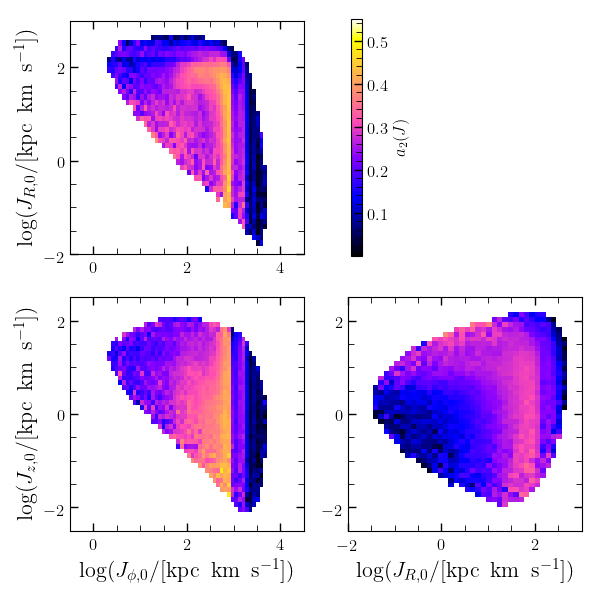}
}
\caption{The strength of the $m=2$ Fourier moment, $a_2(J)$, in model 2 as 
a function of the initial actions.  Only bins with more than 200
particles are shown.  The large values of $a_2(J)$ in the region at
$\log(\act{\phi}/[\kmskpc]) \la 2.8$ are produced by the bar, while
spirals are responsible for the large values at larger \act{\phi}.
\label{f:A2vsJs}}
\end{figure}

We define the $m^{th}$ Fourier amplitude of a population as:
\begin{equation}
a_m(P) = \left| \frac{\sum_{k \in P} m_k e^{i m \phi_k}}{\sum_{k \in P} m_k} \right|,
\end{equation}
where particle $k$ is in the population $P$, and $m_k$ and $\phi_k$
are its mass and azimuthal angle.
Fig. \ref{f:A2vsJs} shows the dependence of $a_2(J)$, the bar strength
of action populations, on the actions.  In the (\act{\phi},\act{R})
plane a large $a_2(J)$ is present at $\log(\act{\phi}/[\kmskpc])
\simeq 2.8$, beyond which is a spiral, which produces the weak
$a_2(J)$ in the outer disc, \ie\ at large \act{\phi}.  In the
$(\act{\phi},\act{z})$ plane $a_2(J)$ appears more uniform across
\act{z}\ at fixed \act{\phi}.  As a result $a_2(J)$ reaches larger values 
in the $(\act{\phi},\act{R})$ plane than in the $(\act{\phi},\act{z})$
plane.  In the $(\act{R},\act{z})$ plane the bar appears as two
branches, one at $\log(\act{z}/[\kmskpc]) \sim 1.5$, and the other at
$\log(\act{R}/[\kmskpc]) \sim 1.7$.  The branch at large \act{z}\ is
comprised of star particles at $R \la 3~\kpc$, while the stars at the
large \act{R}\ branch dominate at larger radii.

\begin{figure}
\centerline{
\includegraphics[angle=0.,width=\hsize]{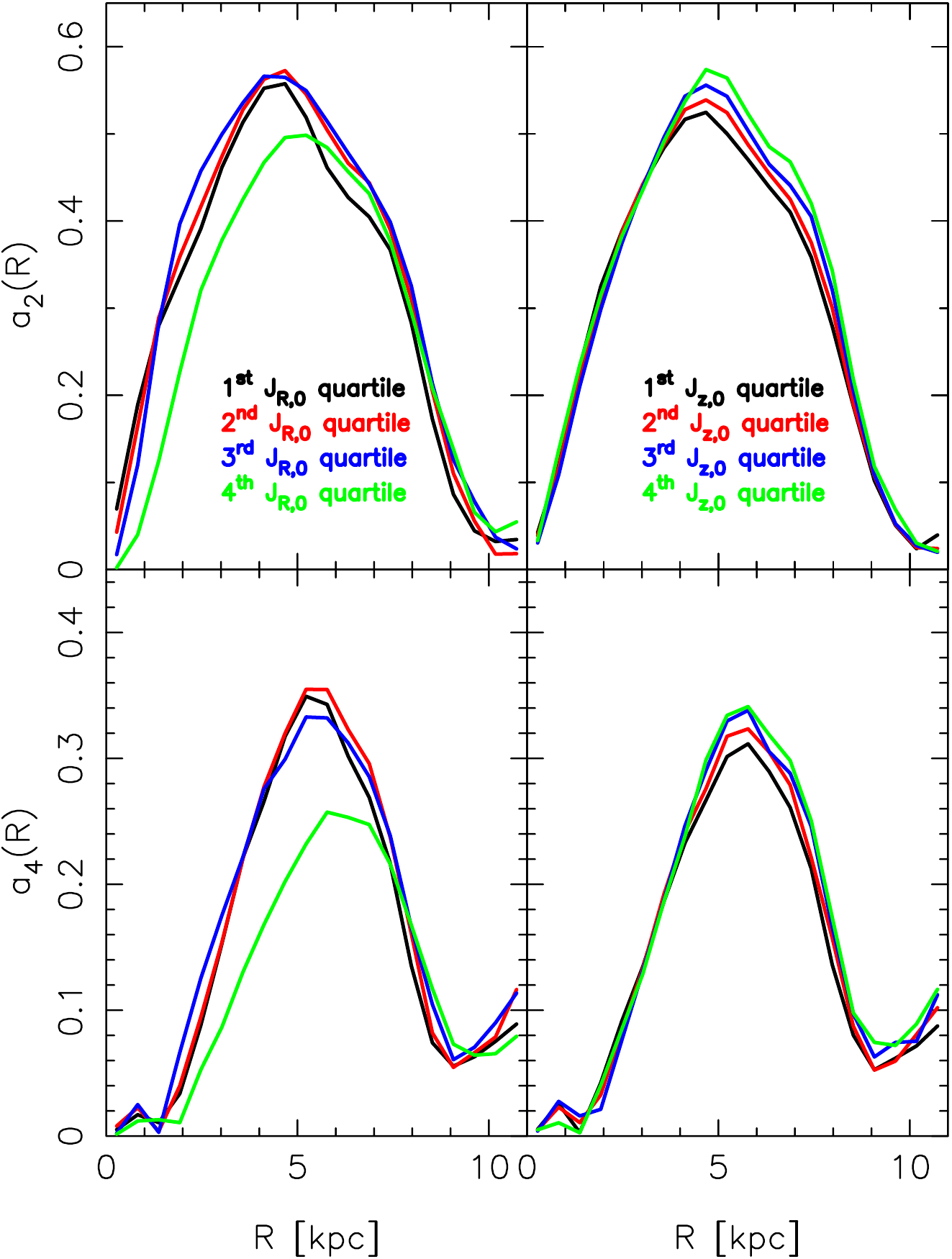}
}
\caption{The bar strength as a function of radius for model 2 split into
different action quartiles.  The top row shows the $m=2$ Fourier
amplitudes while the bottom row shows the $m=4$ amplitudes.  At left
the disc is split into \act{R}\ quartiles while at right into \act{z}\
quartiles.  The bar strength has a larger dynamic range in \act{R}\
than it does in \act{z}.
\label{f:a24vsR}}
\end{figure}

Fig. \ref{f:a24vsR} shows the radial profiles of the $m=2$ and $m=4$
amplitudes, $a_2(R)$ and $a_4(R)$, for the different quartiles.  The
bar is weaker in the highest \act{R}\ quartile.  The differences in
$a_2(R)$ and $a_4(R)$ between different \act{z}\ quartiles are less
pronounced than in the \act{R}\ quartiles.

\subsection{Mean-action maps}
\label{ss:actionmaps}

\begin{figure*}
\centerline{
\includegraphics[angle=0.,width=0.45\hsize]{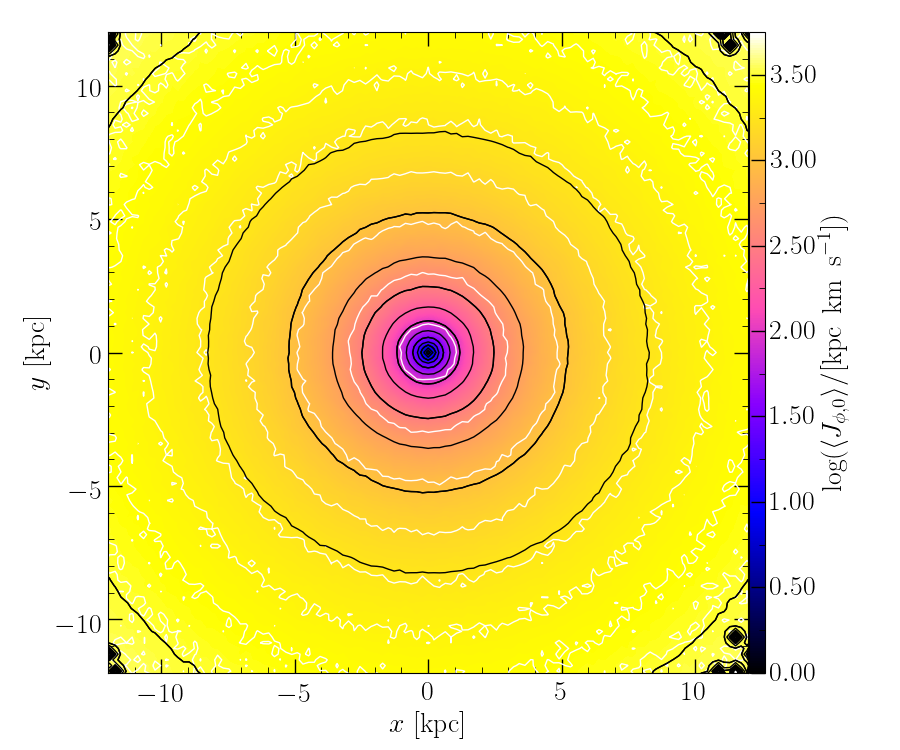}
\includegraphics[angle=0.,width=0.45\hsize]{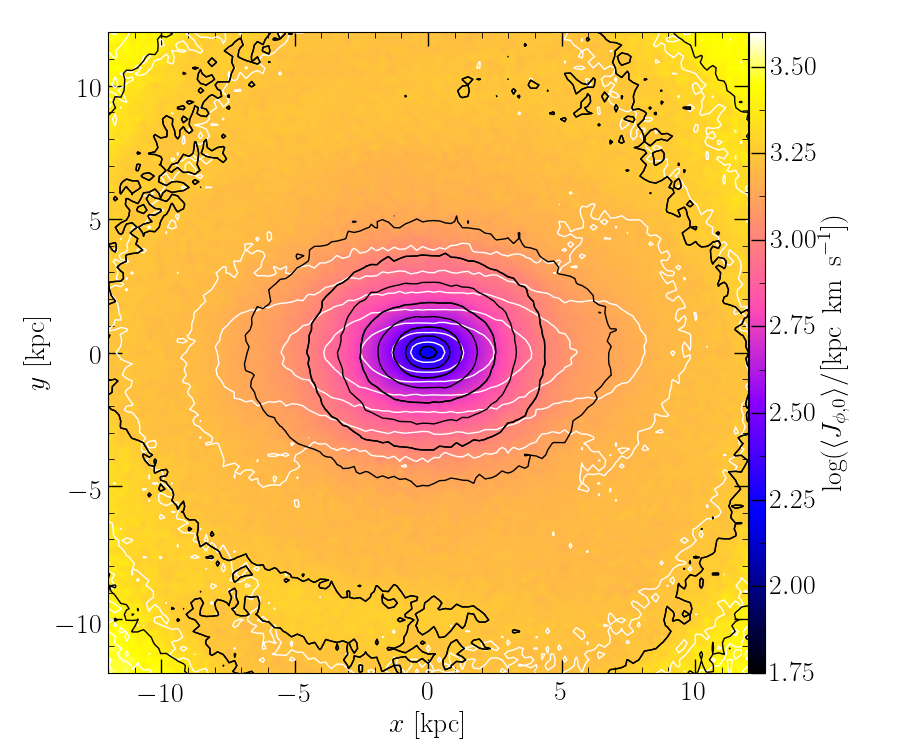}
}
\centerline{
\includegraphics[angle=0.,width=0.45\hsize]{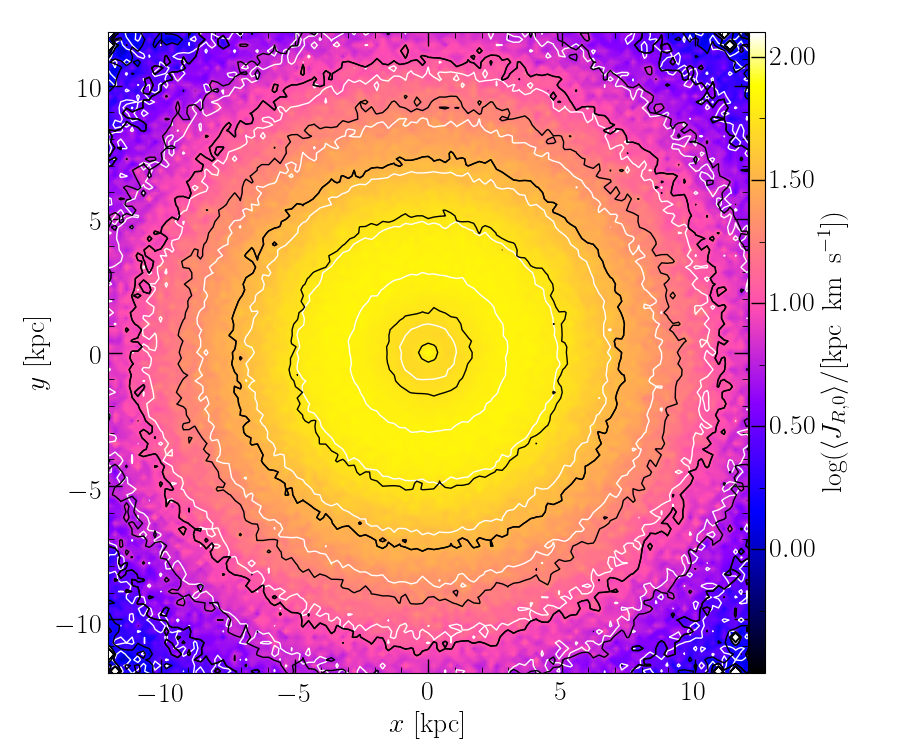}
\includegraphics[angle=0.,width=0.45\hsize]{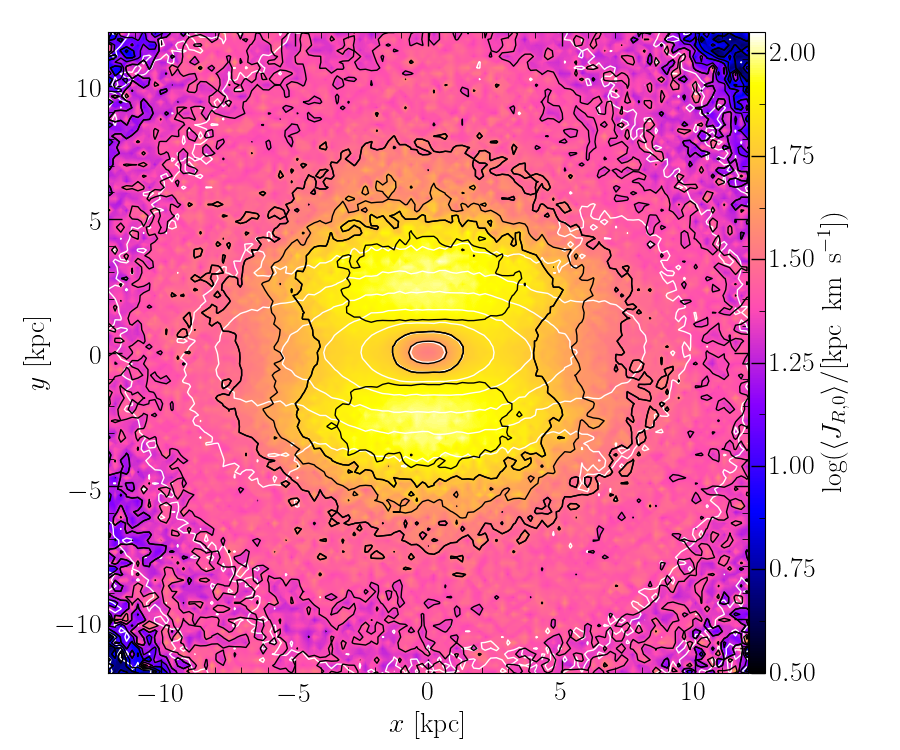}
}
\centerline{
\includegraphics[angle=0.,width=0.45\hsize]{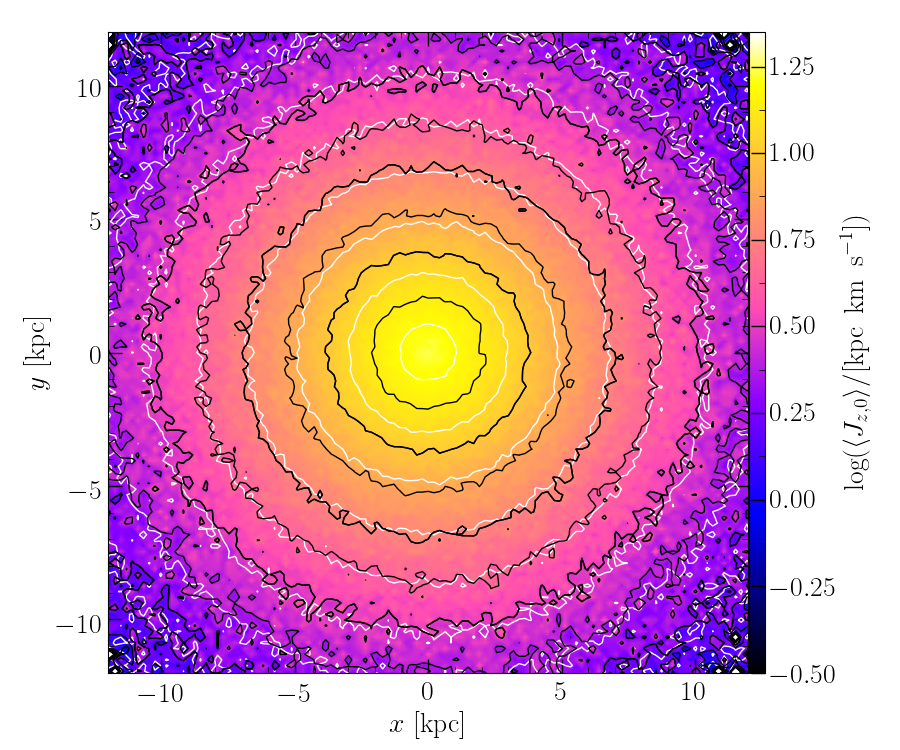}
\includegraphics[angle=0.,width=0.45\hsize]{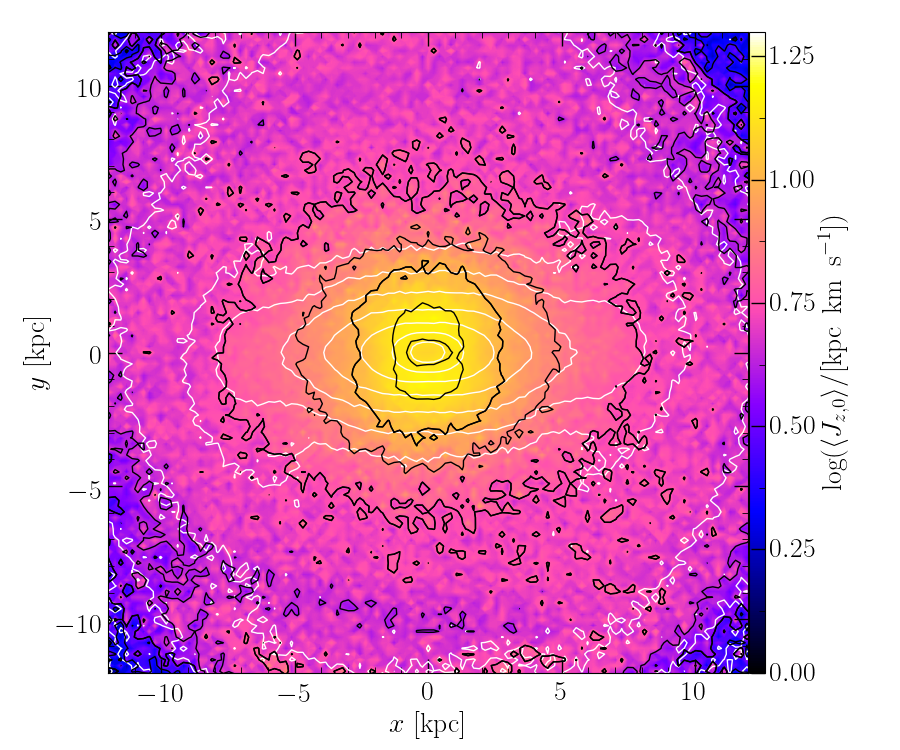}
}
\caption{The face-on surface density (white contours) and mean actions 
(colours and black contours) at $t=0$ (left) and at $t=5\Gyr$ (right)
in model 2.
\label{f:faceon}}
\end{figure*}

Fig. \ref{f:faceon} shows the distributions of mean actions in the
face-on view of the model at $t=0$ and at $t=5\Gyr$.  At $t=0$
\avg{\act{\phi}}\ increases radially outwards, while \avg{\act{z}}\ 
decreases radially.  Instead \avg{\act{R}}\ peaks at $R \simeq
3.8~\kpc$.  At $t=5\Gyr$ \avg{\act{\phi}}\ is elongated along the
major axis of the bar.  In contrast, \avg{\act{R}} and, to a lesser
extent, \avg{\act{z}}, are elongated along the bar's minor axis at
small radii.  These orthogonal orientations are due to the fact that
stars with small \act{R}\ or \act{z}\ are more elongated along the bar
than those with large \act{R}\ or \act{z}\
(Fig. \ref{f:actionquartiles}).  The difference in elongations is
weaker for \act{z}, and therefore the overall elongation along the
$y$-axis is weaker for this action.

\begin{figure*}
\centerline{
\includegraphics[angle=0.,width=0.3\hsize]{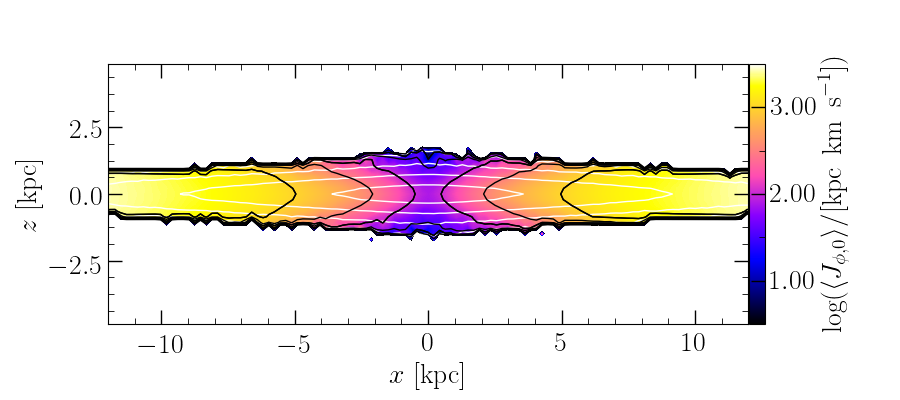}
\includegraphics[angle=0.,width=0.3\hsize]{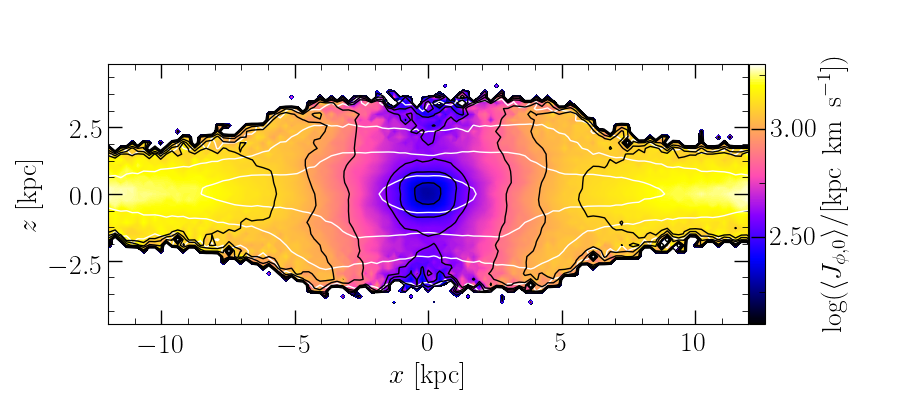}
\includegraphics[angle=0.,width=0.3\hsize]{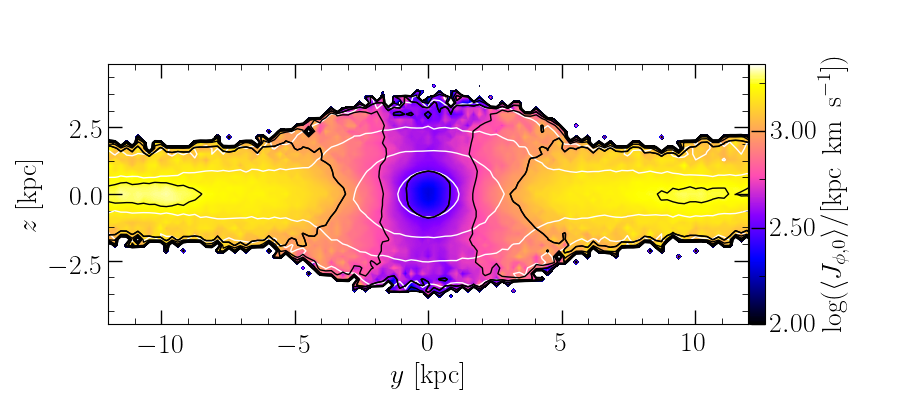}\
}
\centerline{
\includegraphics[angle=0.,width=0.3\hsize]{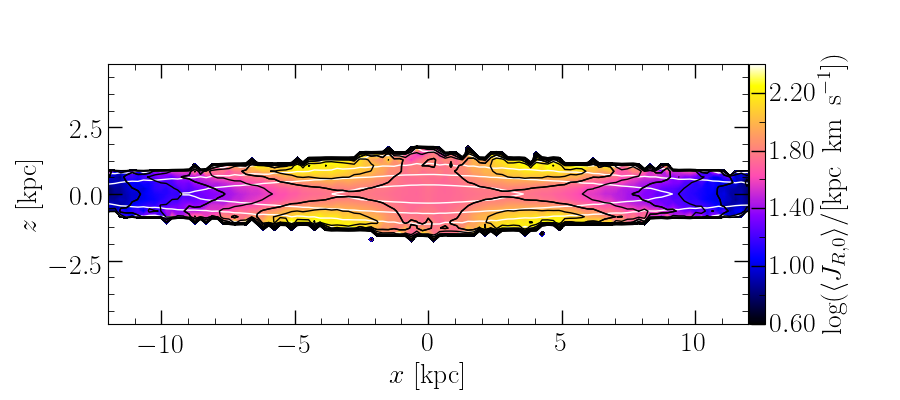}
\includegraphics[angle=0.,width=0.3\hsize]{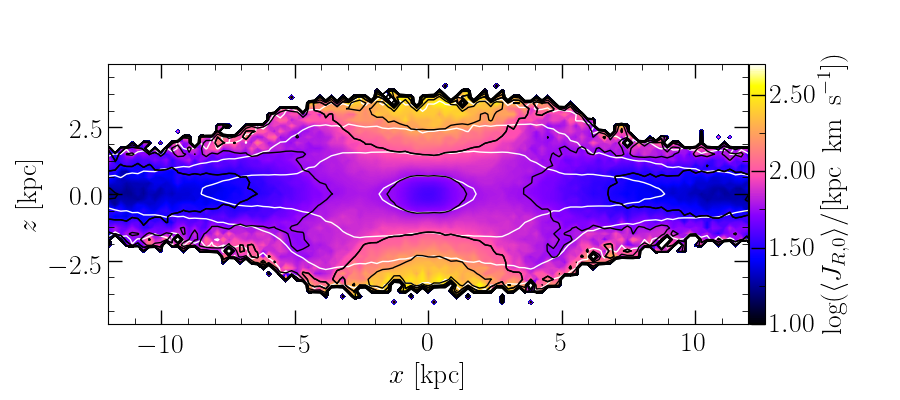}
\includegraphics[angle=0.,width=0.3\hsize]{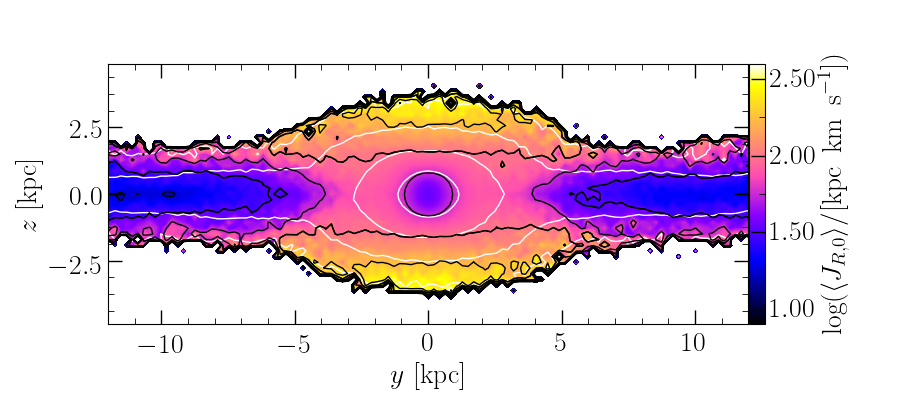}\
}
\centerline{
\includegraphics[angle=0.,width=0.3\hsize]{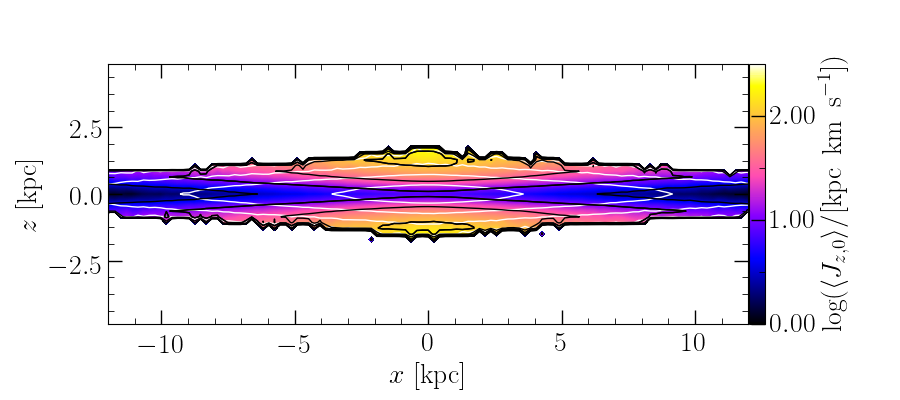}
\includegraphics[angle=0.,width=0.3\hsize]{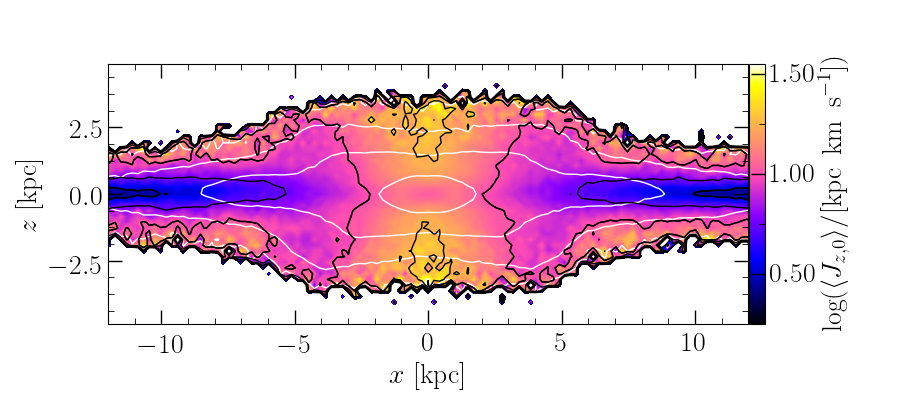}
\includegraphics[angle=0.,width=0.3\hsize]{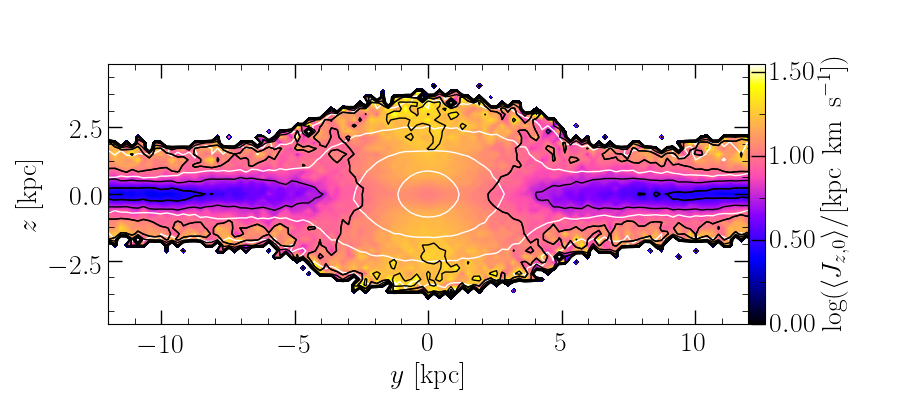}\
}
\caption{Actions in the $(x,z)$ plane at $t=0$ (left) and at the end of 
the simulation, at $t=5\Gyr$ (middle) and in the $(y,z)$ plane (right)
in model 2.  From top to bottom we plot \avg{\act{\phi}},
\avg{\act{R}}\ and \avg{\act{z}}.  Particles are selected in the slice
$|y| < 2\,\kpc$ in the two left columns and in the slice $|x| <
2\,\kpc$ in the right column.  Black contours show the actions, while
white contours show the density.  
\label{f:actions}}
\end{figure*}

Fig. \ref{f:actions} presents maps of the mean actions in the $(x,z)$
plane (\ie\ edge-on) at $t=0$ (left panels) and at $t=5\Gyr$ (\ie\
with the bar viewed side-on, middle panels), and in the $(y,z)$ plane
at $5\Gyr$ (\ie\ with the bar viewed end-on, right panels).  At
$t=0$ \avg{\act{\phi}}\ increases in a radial direction, while
\avg{\act{z}}\ increases in the vertical direction.  The contours of
\avg{\act{R}}\ have a more complicated shape, generally increasing in
the vertical direction and decreasing in the radial direction.  The
peak of \avg{\act{R}}\ is at $R\simeq 3.8~\kpc$ off the mid-plane.  At
the end of the simulation the contours of the \avg{\act{\phi}}\
distribution are centred on the origin and are quite vertical.
Instead, both \avg{\act{R}}\ and \avg{\act{z}}\ have contours that are
pinched when the bar is viewed side-on.  The map of \avg{\act{z}}\ has
a peak on the minor axis but with a quite low gradient, and almost
vertical contours.  The map of \avg{\act{R}}\ also has a peak on the
minor axis but the contours are now more peanut-shaped.  With the bar
seen end-on (right column), the maps of \avg{\act{\phi}}\ and \act{z}\
appear hourglass shaped.  The map of \avg{\act{R}}\ in this projection 
has a significant vertical gradient and exhibits a ``mushroom cap''
shape of high \avg{\act{R}}\ at large $|z|$.

\subsection{Vertical profiles in the bulge}
\label{ss:gradients}

\begin{figure}
\includegraphics[angle=0.,width=\hsize]{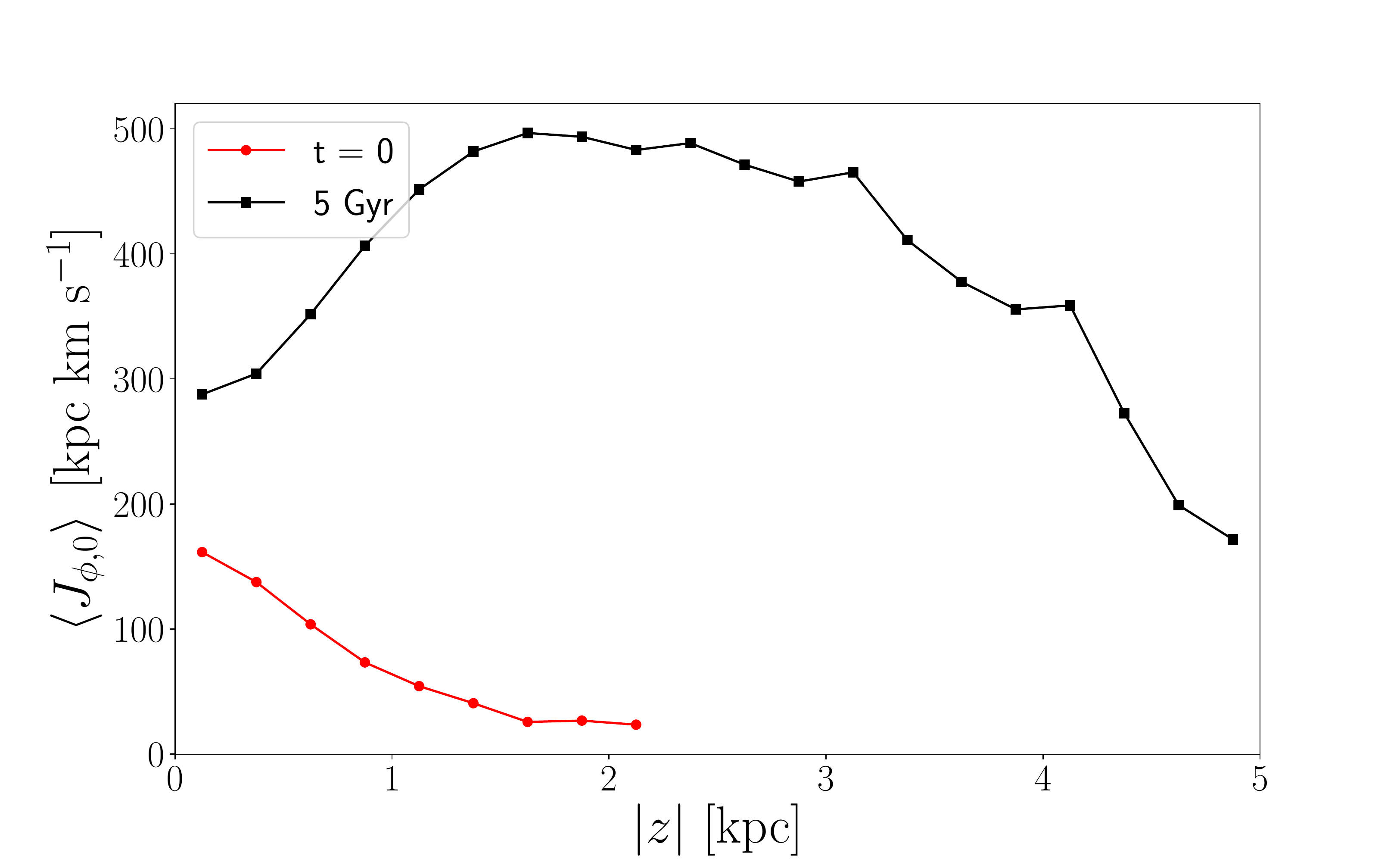}\
\includegraphics[angle=0.,width=\hsize]{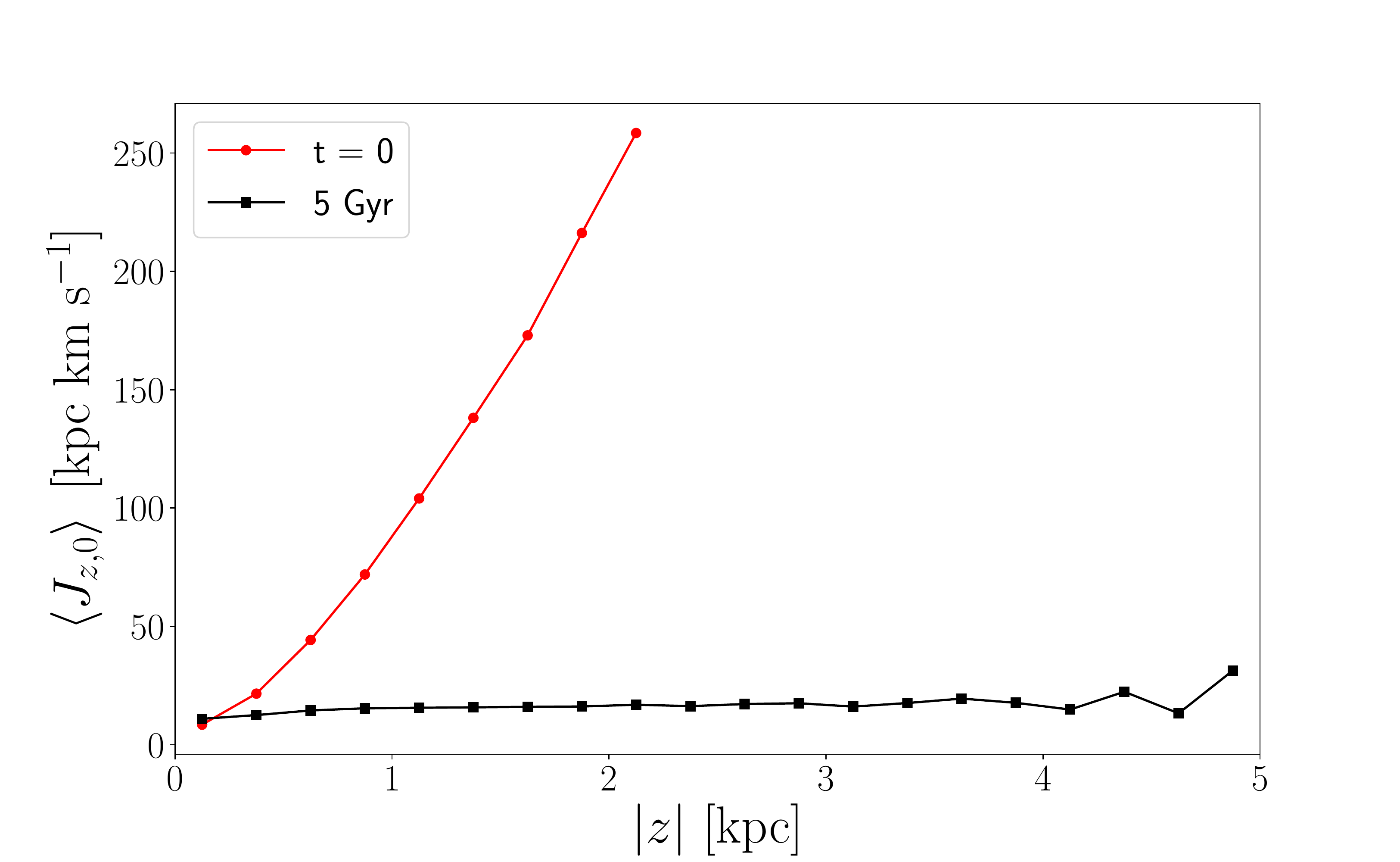}\
\includegraphics[angle=0.,width=\hsize]{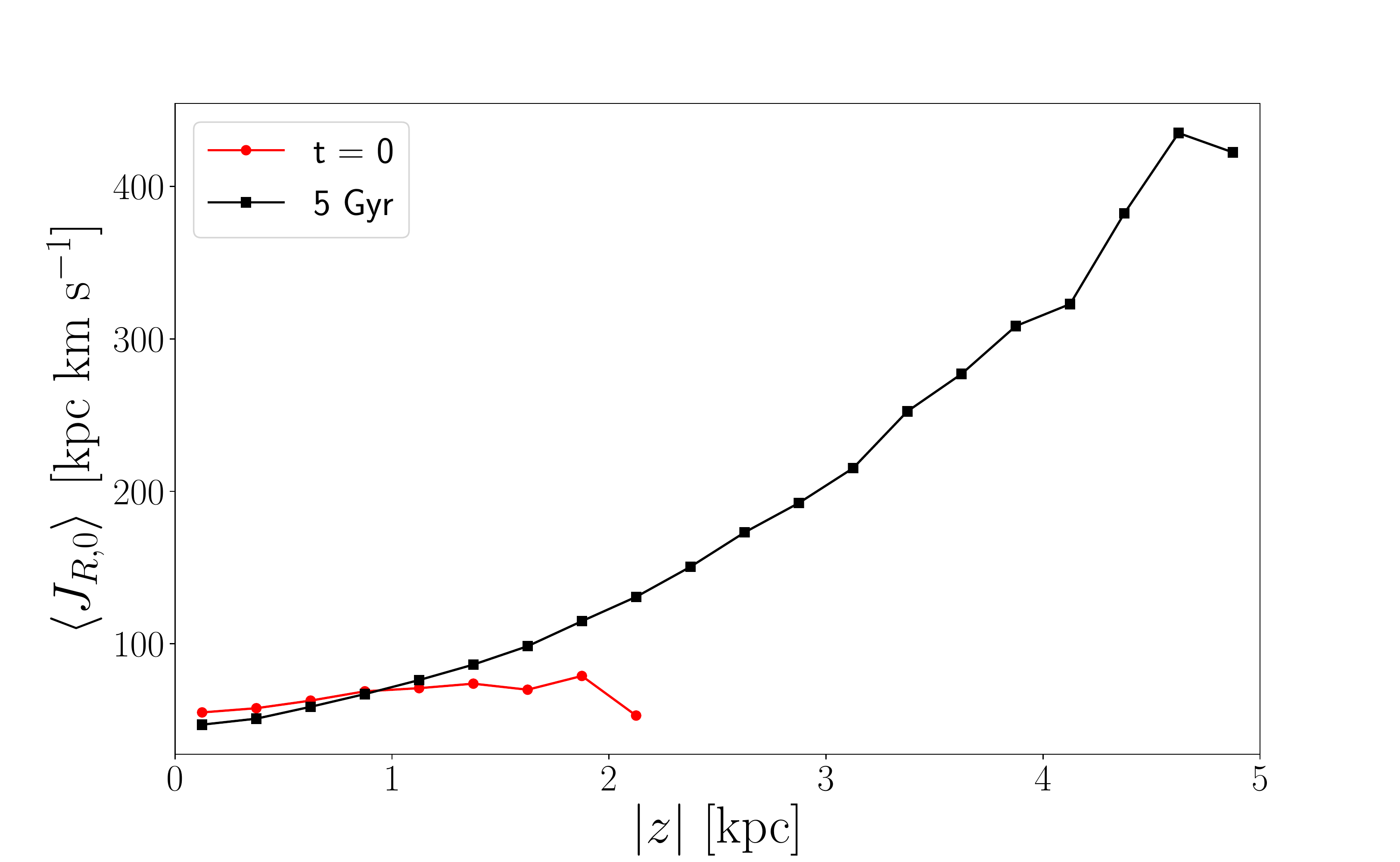}\
\includegraphics[angle=0.,width=\hsize]{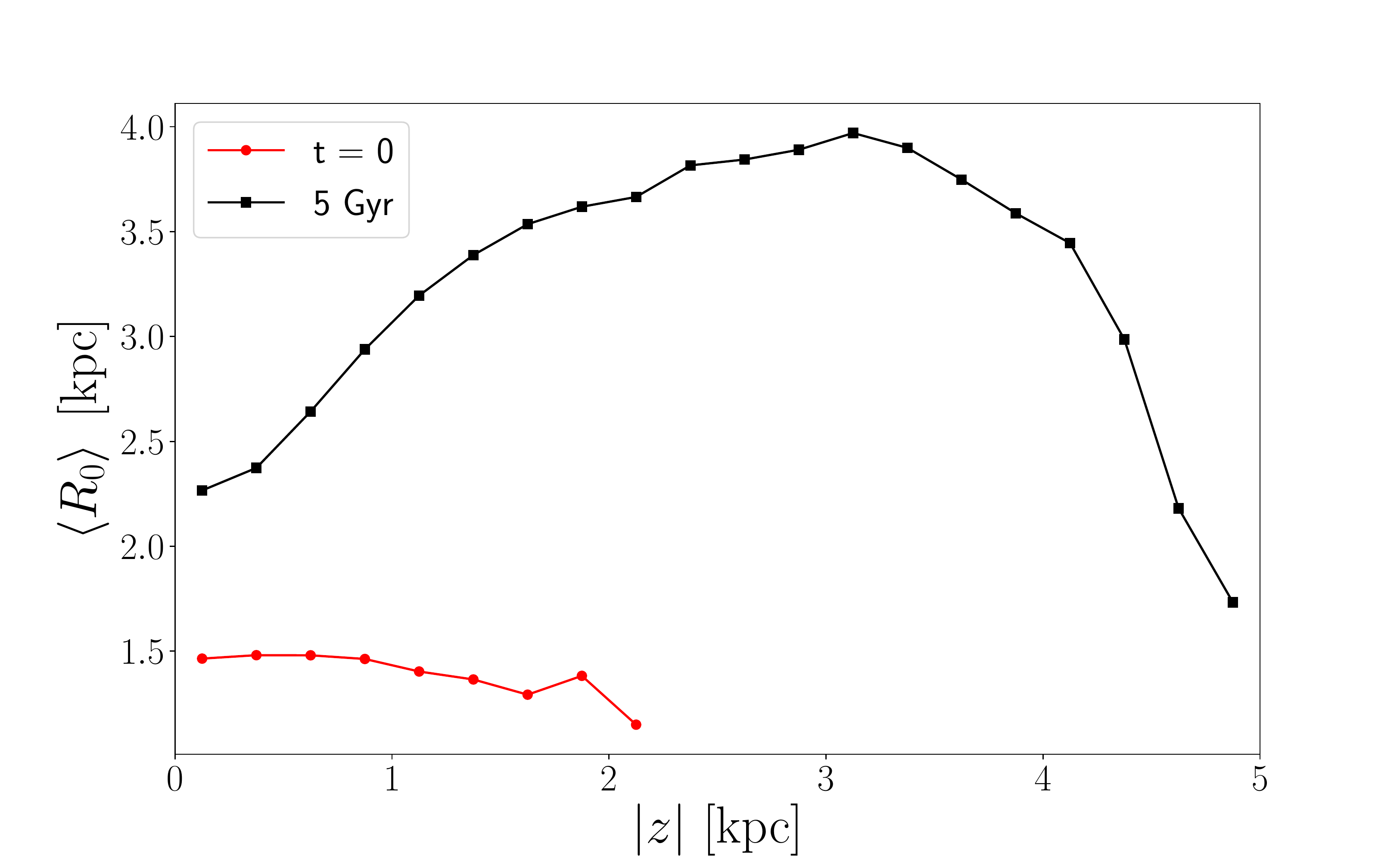}\
\caption{The change in the vertical profiles of \avg{\act{\phi}}\ 
(top row), \avg{\act{z}}\ (second row), \avg{\act{R}}\ (third row),
and \avg{R_0}\ (fourth row) in model 2.  The initial conditions are
shown as (red) filled circles while the final distribution is shown as
(black) filled squares.  All quantities have been computed within the
inner $\Rd = 2.4~\kpc$.
\label{f:vprofiles}}
\end{figure}

\begin{figure}
\includegraphics[angle=0.,width=\hsize]{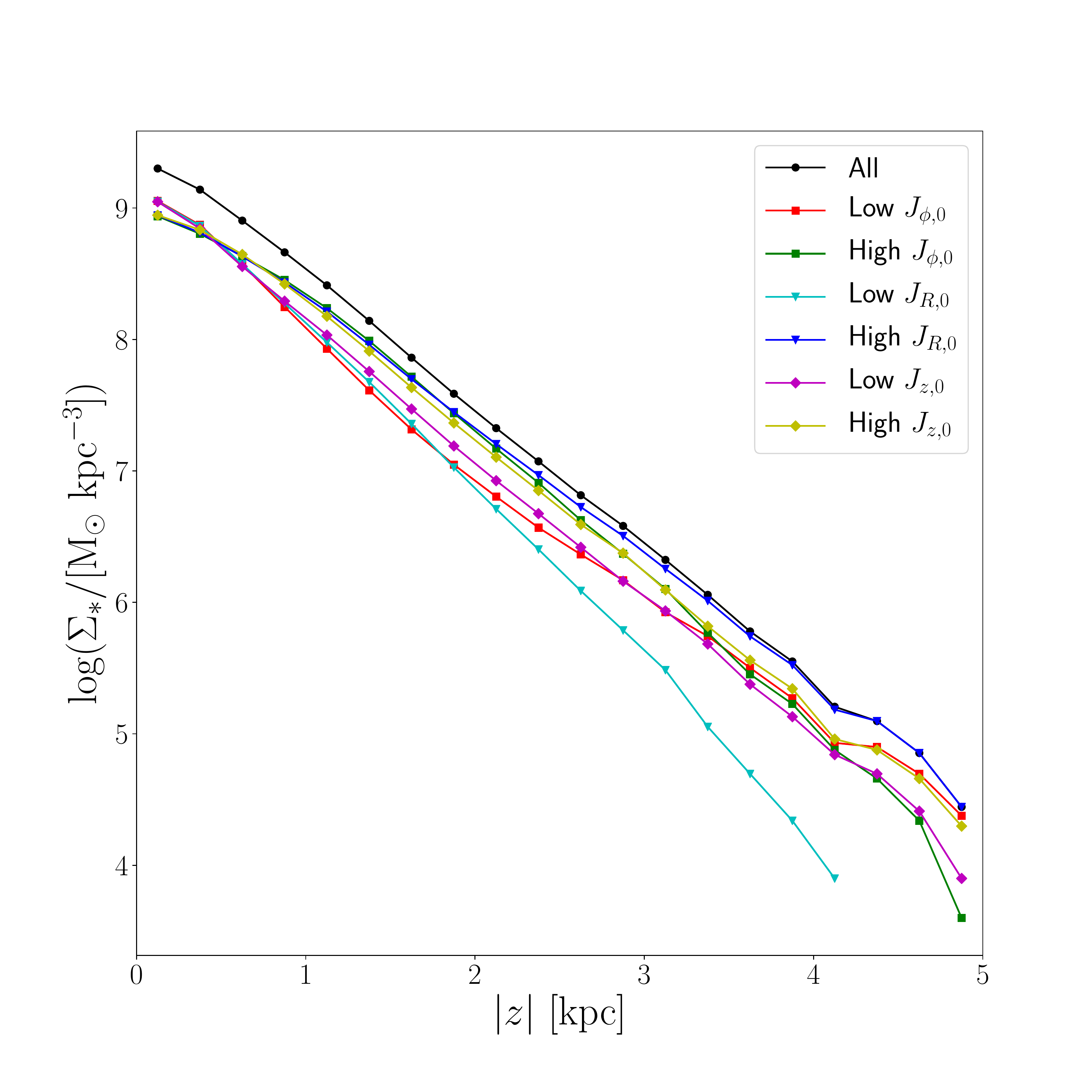}\
\caption{The central vertical density profile of stars in model 2 at 
$t=5\Gyr$ split into two equal samples of low and high actions, as
indicated.  The full sample of stars is shown by the black line.  The
profiles have been computed within the inner $\Rd = 2.4~\kpc$.
\label{f:vprofiles2}}
\end{figure}

Fig. \ref{f:actions} showed that the formation of the B/P bulge
changes the vertical gradients of the actions in the inner galaxy.
Fig. \ref{f:vprofiles} explores the evolution of the vertical profiles
of the actions in the bulge region, defined as inside $\Rd = 2.4~\kpc$
(we have confirmed that the trends shown here are similar inside at
least $3\Rd$, but this is too large a radius for the B/P bulge of some
of the other models, such as model 1).  The initially vertically
declining profile of \avg{\act{\phi}}\ is transformed into a rapidly
rising profile, which peaks at $|z| \simeq 2~\kpc$, and then declines
slowly beyond.  The initially monotonically increasing \avg{\act{z}}\
profile is almost completely flattened by the formation of the B/P
bulge. Conversely the initially rather flat profile of \avg{\act{R}}\
is transformed by the bar into a monotonically rising profile,
indicating that stars with the largest radial motions are the ones
that rise to the largest heights.  Fig. \ref{f:vprofiles2} shows the
vertical density profile for all stars inside $2.4~\kpc$ (black
points) and for the same stars separated into low and high action
halves.  The low and high action profiles are quite parallel, and
overlap, for \act{\phi}\ and \act{z}.  However the vertical profiles
of the low and high \act{R}\ diverge, with the density of low \act{R}\
stars declining more rapidly than that of the high \act{R}.  Above
$3~\kpc$ essentially all stars are from the high \act{R}\ half.

\subsection{Bimodal (X-shaped) distributions}
\label{ss:bimodalities}

\begin{figure*}
\centerline{
\includegraphics[angle=0.,width=\hsize]{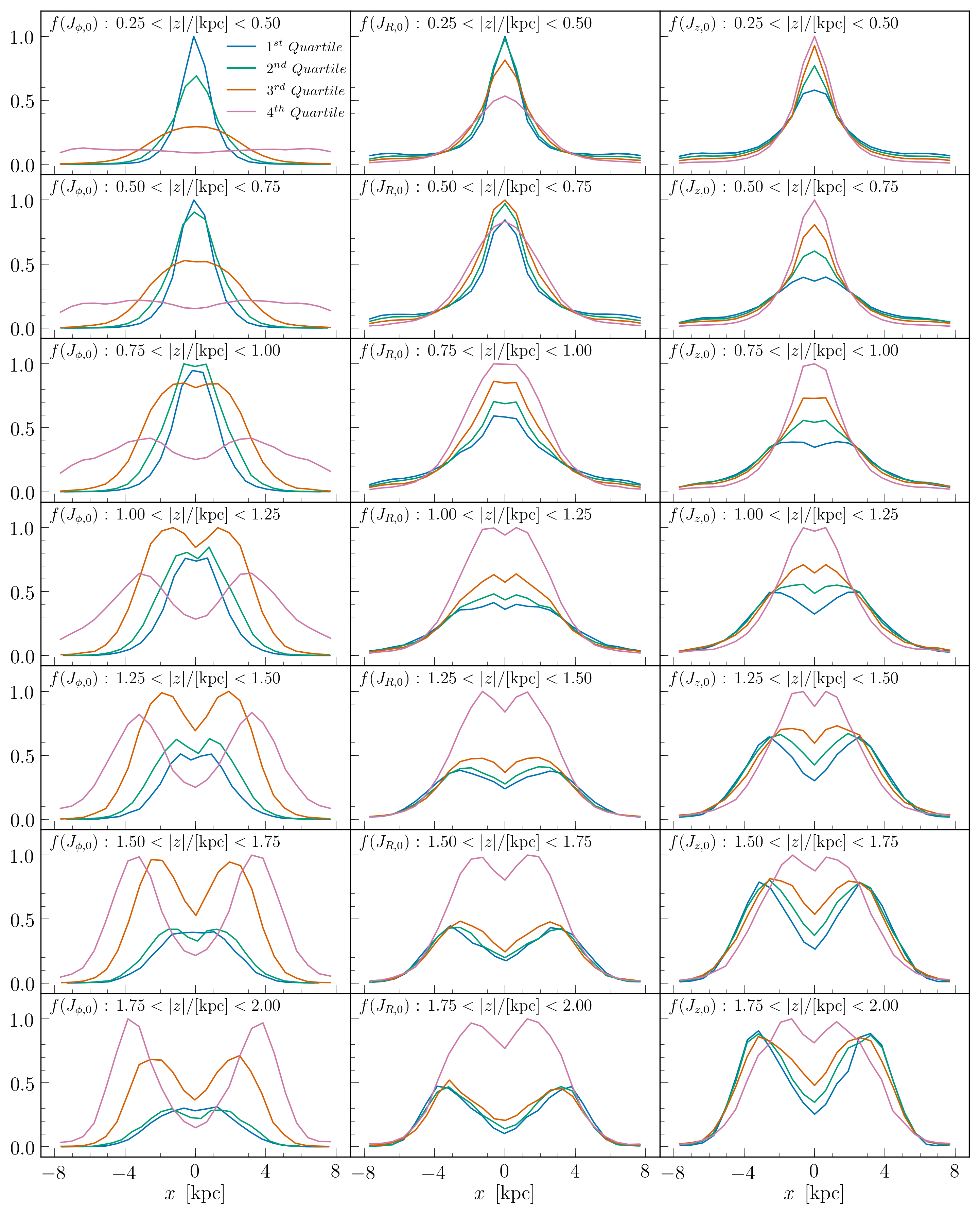} \
}
\caption{Density profiles along the $x$-axis (along which the bar 
is aligned) in model 2 at the end of the simulation, $t=5\Gyr$, for
different heights above the mid-plane. At each height, the profiles
are split into quartiles of the initial actions, as indicated in the
top row, with the $1^{\mathrm st}$ quartile having the lowest
values. Only particles at $|y| < 2 \,\kpc$ are included.  From left to
right are shown the azimuthal, radial and vertical actions.
\label{f:bimodalityJs}}
\end{figure*}

Fig. \ref{f:bimodalityJs} shows the density distribution of stars at
different heights for stars separated into the four quartiles of
\act{\phi}\ (left), \act{R}\ (centre) and \act{z}\ (right) listed in 
Table \ref{t:quartiles}.  All exhibit bimodal distributions for at
least some quartiles.  The separations between the peaks at fixed
$|z|$ increase with \act{\phi}\ and decrease with increasing \act{R}\
and \act{z}.  Moreover these separations increase with larger $|z|$,
\ie\ the distributions are X-shaped.  The behaviour of the density
distributions separated by \act{R}\ and \act{z}\ are somewhat similar.
The separation appears first in \act{\phi}, followed by \act{z}, while
\act{R}\ is the last to show the bimodality.  Of the three actions,
the separation between the peaks is largest in \act{\phi}.

\subsection{Vertical heating and radial cooling}
\label{ss:heating}

\begin{figure*}
\centerline{
\includegraphics[angle=0.,width=0.5\hsize]{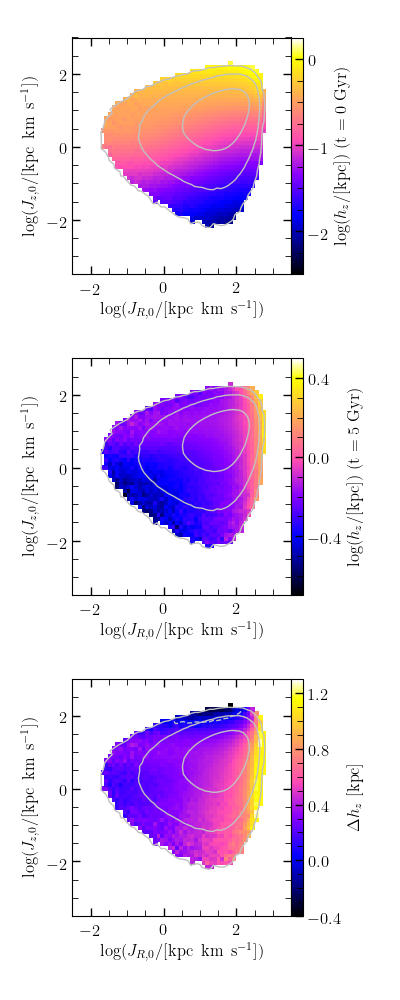}
\includegraphics[angle=0.,width=0.5\hsize]{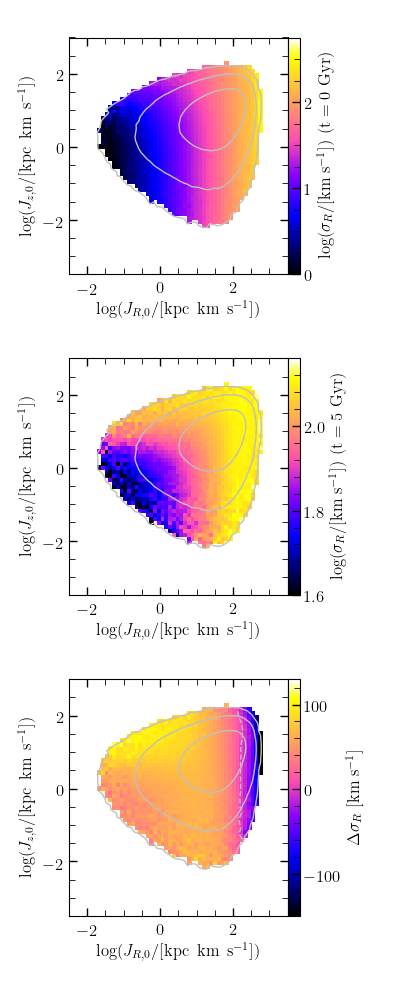}
}
\caption{Left: Evolution of $h_z$, the thickness of stellar populations, 
in the $(\act{R},\act{z})$ plane of model 2.  Right: Evolution of
\sig{R}, the radial velocity dispersion of stellar populations, in the
$(\act{R},\act{z})$ plane of model 2.  The top row shows $t=0$ and the
middle row shows the end of the simulation, $t=5\Gyr$, both on a log
scale.  The bottom row shows the difference, now on a linear scale.
At large \act{z}\ $\Delta h_z < 0$ \ie\ the stellar populations become
thinner after the B/P bulge forms (bottom left).  Similarly, at large
\act{R}, $\Delta \sigma_R < 0$, \ie\ the stellar populations become
radially cooler (bottom right).  The dashed grey contours indicate
$\Delta h_z = 0$ (bottom left) and $\Delta \sig{R} = 0$ (bottom
right).  Solid contours indicate the density of stars and are spaced
by factors of 10, with the peak set at 10,000 particles.
\label{f:acthz}}
\end{figure*}

We define the thickness of stellar populations, $h_z$, as the standard
deviation of the vertical positions of stars.  The left panels of
Fig. \ref{f:acthz} show the evolution of the thickness in
$(\act{R},\act{z})$ space.  At $t=0$ the thickness is a strong
function of \act{z}, which merely shows that the actions are
calculated properly.  The contours of constant $h_z$ are not
horizontal, as a consequence of the radial dependence of the
potential.  After the B/P bulge forms, at $5~\Gyr$, the thickness
increases monotonically with \act{R}, with the thickest populations
those at largest \act{R}.  The thickness increases with
\act{z}\ up to a point, but then decreases again, with the peak
reached never as high as that in \act{R}.  The map of the change in
thickness between the final and initial states, $\Delta h_z$, shows
that $\Delta h_z$ increases most strongly with increasing \act{R}.
Surprisingly vertical heating is negative, \ie\ stellar populations
become thinner, at the largest \act{z}.  Vertical heating therefore is
primarily determined by \act{R}\ and substantially less by \act{z}.  A
comparable figure for the vertical velocity dispersion, \sig{z}, not
shown here, is very similar to the left column of Fig. \ref{f:acthz},
including having a population with $\Delta\sig{z} < 0$.

Conversely we expect \sig{R}\ should also decrease in the populations
with large \act{R}.  We test this prediction in the right panels of
Fig. \ref{f:acthz}, which show that although most of the system heats
radially, all populations at $\log(\act{R}/[\kmskpc]) \ga 2.2$ cool.
The contour with $\Delta\sig{R} = 0$, shown by the dashed grey line,
is surprisingly vertical considering that the bar instability
increases the radial random motion, driving \sig{R}\ up, and setting up
the radial anisotropy which drives the buckling instability
\citep{araki_phd, merritt_sellwood94}.  The fraction of the disc that
radially cools is larger than that which cools vertically.


\section{Different initial conditions}
\label{s:simsuite}

The analysis thus far has considered only the fiducial model 2.  Now
we consolidate these results by considering the rest of the simulation
suite.  The appendices present figures similar to Figs. \ref{f:model}
to \ref{f:acthz} for the rest of the models.  This section discusses
trends that arise from varying the radial random motion, varying the
thickness, from thin$+$thick disc structure and from a high dark
matter fraction.  Section \ref{ss:synthesis} provides a synthesis of
the simulation suite.

\subsection{Varying the radial velocity dispersion}
\label{ss:radialdisp}

Models 1 and 3 are initially radially cooler and hotter, respectively,
than the fiducial model 2.  Model 1 has the narrowest range of
\act{R}\ of any of the baryon-dominated models (the model has
Toomre-$Q \sim 0.8$ over a short radial range).  The bar strength
barely varies with \act{R}\ or \act{z}, and this is the only
baryon-dominated model in which the vertical gradient is more
prominent in \act{\phi}\ than in \act{R}, reflecting on the quite
narrow range of \act{R}.  In the inner $\Rd$, the separation of the
vertical profiles of the action halves is larger in \act{\phi}\ than
in \act{R}. In model 3 the bar amplitude depends strongly on \act{R}\
and less so on \act{z}.  In the face-on distributions, the peaks in
\avg{\act{R}}\ on the minor axis of the bar become more prominent as
the radial random motion increases.  In the side-on view, pinching of
\avg{\act{R}}\ is present in all three models.  In model 3 the arms of
the X-shape become evident at larger height for both \act{R}\ and
\act{z}.  In all three models the separation of peaks in the
bimodality is strongest in \act{\phi}, with the highest quartile
always being the most widely separated.  All three models are
thickest, and thicken the most, at the largest \act{R}, including in
model 1 which has the narrowest range of \act{R}.  In contrast
populations with the largest \act{z}\ always get thinner after the B/P
bulge develops.  In model 3, the vertical profile of \avg{\act{R}}\ is
still rising at $|z|=5~\kpc$, and the vertical density of the low
\act{R}\ half declines rapidly, while that of the high \act{R}\ half
dominates (over all action halves) from $|z| \simeq 2~\kpc$.

\subsection{Varying the thickness}
\label{ss:thickness}

Models 4 and 5 have thinner and thicker initial discs than the fiducial
model, respectively.  Model 4 thickens considerably, in spite of
having one of the mildest bucklings.  Its initially low thickness
ensures that the strong bar that forms produces a highly anisotropic
velocity distribution, leading to a strong vertical heating.  Because
it starts out so thin, the region of action space where the
populations become thinner is reduced compared to the fiducial model,
but it is still present.  The final vertical profiles of the actions
follow the same trend as in the fiducial model.  All four \act{z}\
quartiles have nearly identical bar amplitude, while a large weakening
in the bar strength and peanutiness is evident for the highest
\act{R}\ quartile.  In the initially thickest disc, in model 5, the
highest \act{z}\ populations still end up thinner after B/P formation
over a more extended region than in the fiducial model.  The bar
strength and peanutiness is still better separated by \act{R}, rather
than by \act{z}.  The bar amplitudes are similar in all the
populations except for the highest \act{R}\ one in model 4 while in
model 5 the difference is largely in the radius of the peak amplitude
rather than the peak itself.  As in models 1-3, the populations that
thicken the most, and end up thickest, are the ones with the largest
\act{R}.

\subsection{Thin$+$thick disc models}
\label{ss:thinthick}

Model T1 exhibits much the same trends as the fiducial model.  The bar
is strongest in the lowest quartiles of \act{R}\ and \act{z}, with the
difference between bar strengths greater when the model is split by
\act{R}\ than by \act{z}.  Only a very minor population of model T1 ends 
up thinner at the end of the simulation.  On the other hand a
significant population becomes radially cooler.  

\begin{figure}
\centerline{
\includegraphics[angle=0.,width=\hsize]{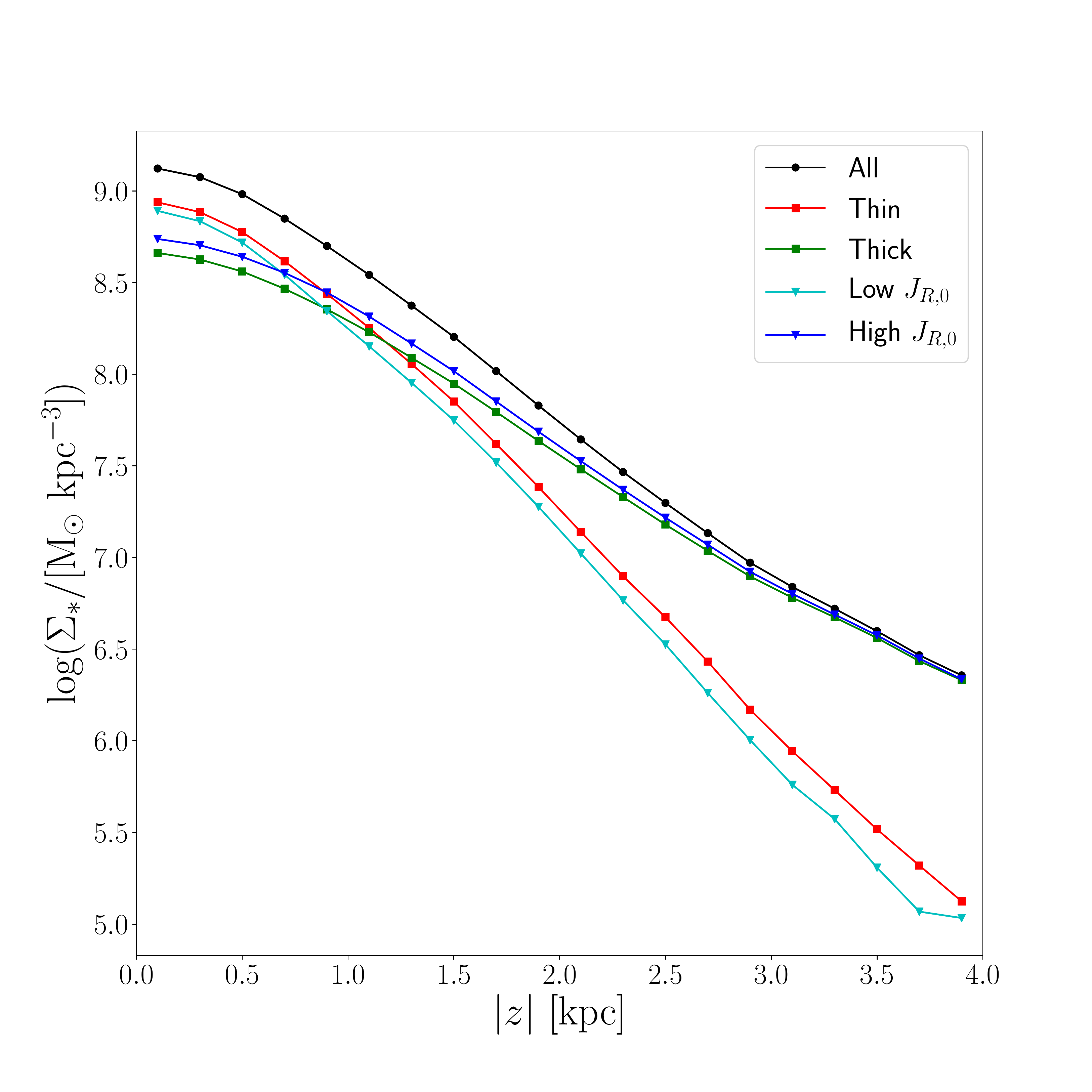}
}
\caption{The vertical density profile of stars within $\Rd = 2.4~\kpc$ 
at $t=5\Gyr$ in model T5.  The full profile (black) is split into
thin and thick discs (red and green respectively) and into low and
high halves of \act{R}\ (cyan and blue).
\label{f:T5vertprof}}
\end{figure}

In model T5 the population with the largest \act{R}\ has a significant
weaker bar than in any other population.  The smaller range of bar
strengths in \act{z}\ quartiles compared to \act{R}\ quartiles is
particularly clear in this case and the highest \act{R}\ population
has $a_4 \simeq 0$. The vertical gradient in \act{z}\ is considerably
weakened by bar and B/P-bulge formation whereas once more an enhanced
gradient in \act{R} results.  Fig.~\ref{f:T5vertprof} shows the final
vertical density profile of model T5 in the inner disc, as well as the
same stars separated into halves of \act{R}, and into those from the
thin and thick discs.  At large heights, stars are more likely to have
originated from the higher \act{R}\ population than they are from the
thick disc.  Even when there is a thin and a thick disc dichotomy to
start with, the population that ends up thickest, and thickens the
most, is still the one with the largest \act{R}.  Nonetheless, a
significant populations cool vertically and radially.

\subsection{Dark-matter dominated systems}
\label{ss:darkmatter}

The properties of model HD1 are similar to those of model 1.  What
these two models have in common is a quite narrow range of
\act{R}\ (see Table \ref{t:quartiles}).  As in model 1, the vertical
profiles of the action halves in the inner galaxy are more separated
in \act{\phi}\ than in \act{R}.  In addition, while the formation of
the B/P bulge retains a vertical gradient in \act{R}, the resulting
profile is shallower than at the start.  Another similarity between
the two models is that the X-shape has comparable strength in all the
action quartiles at large heights from the mid-plane.  Additionally
the bar amplitude is essentially identical in all \act{R}\ and
\act{z}\ quartiles.  The two peaks are clearly separated, regardless
of which action is considered.  While a population that cools
vertically is readily apparent, unlike model 1 only a very
insignificant population cools radially.  Model HD2 has a minimum
Toomre-$Q \sim 1.2$, but it still has a narrower range of \act{R}\
than does model 1.  While in most respects it behaves like model 2,
the different quartiles have very similar bar amplitudes, while the
X-shape is very similar across all the quartiles.  Unlike model HD1,
in model HD2 a steeper vertical gradient in \act{R}\ develops.

\subsection{Synthesis of the simulation suite}
\label{ss:synthesis}

Overall the simulations exhibit a number of important commonalities.
The dominant role of the radial action in predicting the final
distributions of stars in the B/P bulge is attested by the
monotonically increasing vertical heating with increasing \act{R}, the
monotonic vertical profile of \avg{\act{R}}\ within the B/P bulge, and
the larger separation in bar strengths when stellar populations are
separated by \act{R}.  All the models end up thickest in the
populations with the largest \act{R}.  In all models a bimodal
distribution, \ie\ an X-shape, is present at some distance away from
the mid-plane; the separation is largest at the highest \act{\phi}\
and in the lowest \act{R}\ and \act{z}. The vertical density profiles
of the stars in the \act{\phi}\ and \act{z}\ halves are quite
parallel, but diverge in the \act{R}\ halves.  The only models that
diverge significantly from these trends are the ones with a narrow
range of \act{R}, which happens when the initial disc dips below
$Q=1$.


\section{Stellar population modelling}
\label{s:chemistry}

The chemistry of bulges provides an important constraint on their
formation.  There are several observational trends that models of B/P
bulges need to satisfy.  Firstly, in the MW, the density bimodality,
\ie\ the X-shape, is observed to be pronounced in metal-rich stars and
weak or absent in the metal-poor ones \citep{ness+12, uttenthaler+12,
rojas-arriagada+14}.  Secondly, but related to this, self-consistent
chemo-hydrodynamical simulations \citep{debattista+17, debattista+19,
athanassoula+17} have predicted more pinched metallicity distributions
than the density itself when viewed edge-on; this distribution has
been confirmed in real galaxies \citep{gonzalez+17}.  This trend is
the equivalent of the metallicity dependence of the X-shape,
generalised to external galaxies.  Lastly, along the MW's minor
(vertical) axis, a metallicity gradient is observed
\citep{gonzalez+11, gonzalez+13, ness+13a, zoccali+17}.

In order to explore the relation between the action trends found above
and the metallicity trends, we now develop a method for assigning
metallicity to star particles based on their actions.  We employ a
self-consistent star-forming chemo-hydrodynamical simulation to
determine how the metallicity depends on the actions in an unbarred
galaxy.  We then use actions to map metallicity from the star-forming
simulation to the star particles in the fiducial, pure $N$-body,
model.  Since the metallicity of stars in real galaxies depends, if
anything, on the actions \emph{before} the bar forms, our emphasis
here on the initial actions is well-matched to the problem at hand.
First, however, we consider the limitations of modelling the
metallicity based on only single actions to motivate full three-action
modelling.

\subsection{Metallicities based on single actions}

\subsubsection{Metallicity based on \act{\phi}\ or initial radius}

\citet{martinez-valpuesta_gerhard13} investigated the origin of the MW's 
vertical metallicity gradient in pure $N$-body simulations by assuming
that the initial axisymmetric disc had a steep radial metallicity
gradient.  They showed that in this case, a metallicity gradient
similar to the one observed arises.  Consistent with this result, the
vertical profile of the mean initial radius, \avg{R_0}, shown in the
bottom panel of Fig. \ref{f:vprofiles}, has a shallow declining
profile at $t=0$, which is transformed into a rising profile, with a
decreasing gradient.  The same happens in the full simulation suite
but in many cases the profile of \avg{R_0}\ reaches a peak and then
declines again.  These profiles are very similar to the profiles of
\act{\phi}, and assigning metallicity by radius can be considered 
comparable to assignment by the angular momentum.

Assuming metallicity that declines with increasing
\act{\phi}\ results in an X-shape that is stronger in metal-poor
stars than in metal-rich ones. Liu et al. (in progress) reach a
similar conclusion.  This can be seen in Fig. \ref{f:bimodalityJs},
which shows that it is the populations with largest \act{\phi}\ which
are most widely separated, which is opposite to the trend observed in
the MW.  Assigning metallicity by \act{\phi}, or nearly equivalently,
by radius, therefore produces the wrong X-shape trends compared to
those observed in the MW.

\subsubsection{Metallicity based on \act{z}\ or initial height}

Assigning metallicity by \act{z}\ is similar to assuming an initial
vertical metallicity gradient (although \act{z}\ is a better measure
of a population's thickness than the instantaneous height).  Assuming
metallicity depends on \act{z}\ produces an X-shape which is better
traced by metal-rich stars, and metallicity distributions that are
pinched.  For the entire simulation suite we observe that the initial
vertical gradient in \act{z}\ is substantially flattened by the time
the B/P bulge forms.  Any vertical metallicity gradients are therefore
substantially weakened and unlikely to be preserved.  An observational
test of metallicity depending primarily on \act{z}\ is that unbarred
galaxies should have steeper vertical gradients than barred ones.
Since this appears not to be the case \citep{molaeinezhad+17}, we
conclude that \act{z}\ (or height) is not a reliable way of assigning
metallicities to particles.

\subsubsection{Metallicity based on \act{R}\ or age}

\act{R}\ measures the radial random motion; stars are generally born on 
nearly circular orbits and slowly heat radially.  Therefore \act{R}\
might be used as a proxy for age, particularly for thin discs.
Assuming older stars are more metal-poor would allow metallicity to be
assigned based on \act{R}.  This produces most of the trends observed
in B/P bulges, including a stronger X-shape in metal-rich stars, a
vertical metallicity gradient, which can be stronger than in unbarred
galaxies, and a pinched metallicity distribution when observed
edge-on.  The vertical metallicity gradient arises naturally via the
monotonically rising profile of \act{R}\ produced by the
buckling. \citet{molaeinezhad+17} find some evidence for a steeper
metallicity gradient in barred galaxies than in unbarred galaxies.

The principal drawback of using \act{R}\ for metallicity modelling is
that a single value of \act{R}\ corresponds to a wide range of
\act{\phi}, with often very different bar strengths (see Fig. 
\ref{f:A2vsJs}).  Moreover, in our  simulation suite, \act{R}\ does not 
peak at the centre of the galaxy at $t=0$ (see Fig. \ref{f:faceon}),
which would require that the metallicity peaks away from the centre.
Furthermore, the peak of \act{R}\ is generally off the mid-plane (see
Fig. \ref{f:actions}).  Therefore a metallicity based only on \act{R}\
is unlikely to be realistic.

\subsubsection{Metallicity based on radial and vertical position}

\citet{bekki_tsujimoto11} used the radial and vertical position of a star 
particle in the initial axisymmetry system to assign metallicity.
Using the prescription of \citet{bekki_tsujimoto11}, we find that the
X-shape has comparable strength in each metallicity population,
contrary to what is observed in the MW.  Fundamentally the problem is
that all star particles at a given position are treated the same when
they should have different ages, and therefore different \act{R}\ and
different metallicities.

We therefore conclude that assigning metallicities based on actions
requires all three actions.  Below we develop a method for such
modelling.

\subsection{A star-forming model}

\begin{figure}
\centerline{
\includegraphics[angle=0.,width=\hsize]{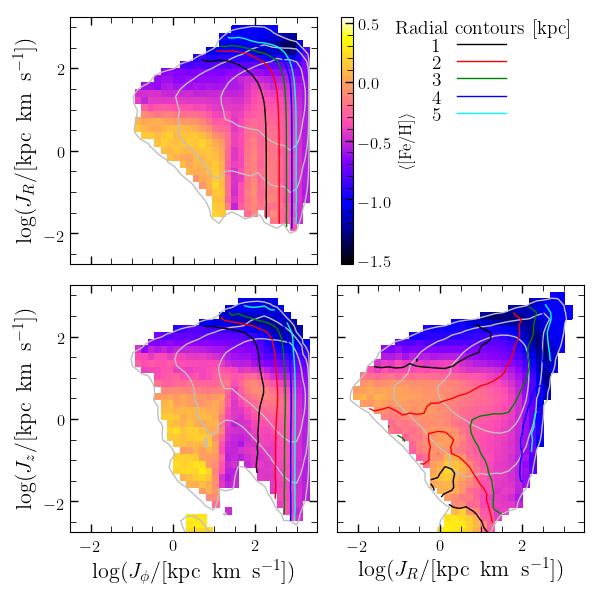}
}
\caption{The mean metallicity of the star-forming simulation at 
$3\Gyr$ in action space.  Coloured contours show the mean radii of
particles, as indicated.  Solid grey contours indicate the density of
stars and are spaced by factors of 10, with the peak set at 10,000
particles.  Only bins with more than 100 particles are shown.
\label{f:sfsim}}
\end{figure}

The star-forming model we use to connect actions and metallicities is
a generic example of the many MW-mass models we have constructed and
presented in the past \citep[e.g.][]{roskar+08a, loebman+11,
loebman+16} but with the $5\times$ higher resolution used in
\citet{portaluri+17}.  It evolves self-consistently, with all stars
forming out of gas cooling off a hot corona.  We use a slightly higher
supernova feedback efficiency than in those earlier works, at $4\times
10^{50}$ erg per supernova.  All other parameters are identical to the
earlier works and will not be described here, but we anticipate
describing the evolution of this simulation in detail elsewhere.

We use this simulation strictly to query the chemistry as a function
of the actions.  We consider the model at $t=3\Gyr$, on the premise
that this is a plausible time for when the bar of the MW formed.
Fig. \ref{f:sfsim} shows the density and mean metallicity of the
star-forming model in the action space at this time.  The density
distribution in the $(\ac{R},\ac{z})$ plane is qualitatively similar
to that in the $(\act{R},\act{z})$ plane of the fiducial model
(Fig. \ref{f:acthz}).

\subsection{Three-action assignment of the metallicity}
\label{ss:tagging}

\begin{figure}
\includegraphics[angle=0.,width=\hsize]{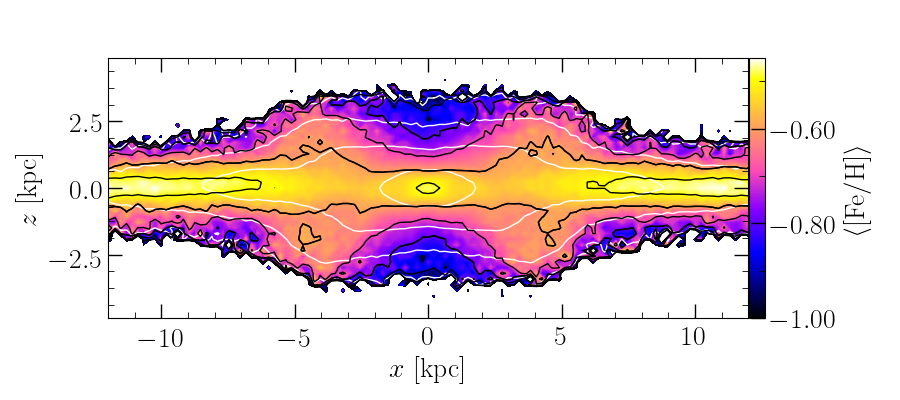}
\includegraphics[angle=0.,width=\hsize]{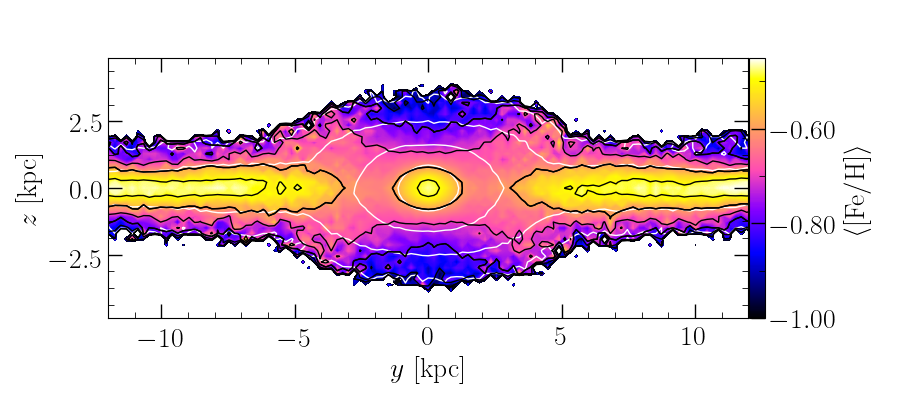}
\includegraphics[angle=0.,width=\hsize]{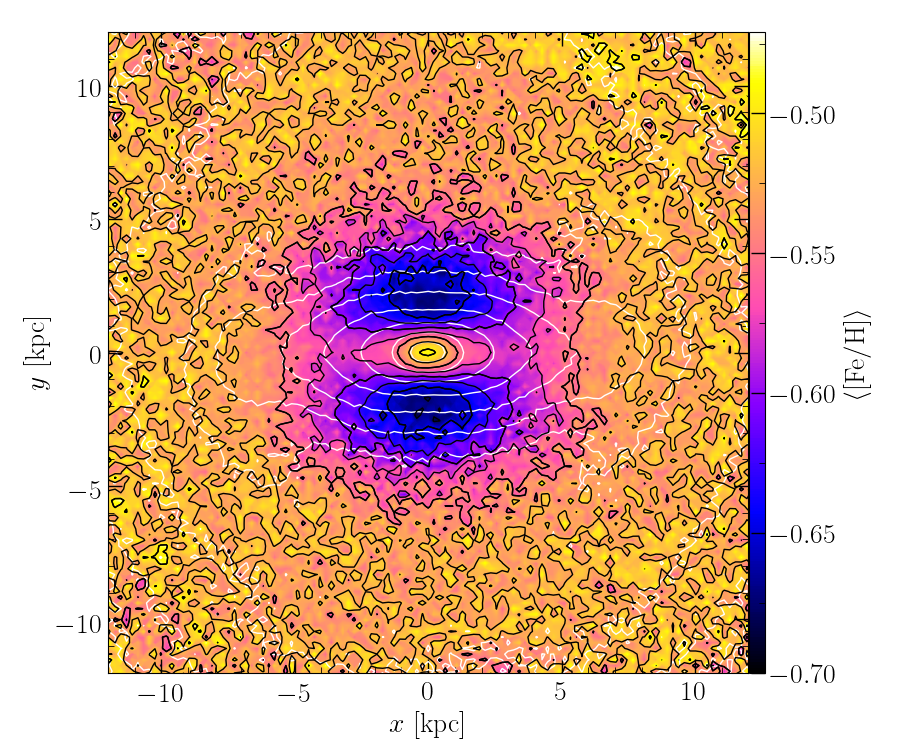}
\caption{Maps of the mean metallicity of model 2 based on metals mapped 
using the three actions as described in Section \ref{ss:tagging}.
Top: Edge-on view with the bar seen side-on.  Middle: Edge-on view
with the bar seen end-on.  Bottom: Face-on view.  Black contours show
the \avg{\feh}\ while white contours show the density.
\label{f:metalmaps}}
\end{figure}

We now demonstrate assigning metallicities to star particles in the
fiducial $N$-body model 2 (hereafter ``the target'') using the
metallicities of star particles in the star-forming model (hereafter
``the donor'').  Because the density distribution, and the actions, in
these two models are quite different, we cannot assign metallicities
by directly matching the actions.  We start by cutting the donor
to only those particles within the volume $|x|,|y| \leq 8~\kpc$ and
$|z| \leq 2~\kpc$.  We then construct the cumulative distribution
functions, $F_{d,\phi}$, $F_{d,R}$ and $F_{d,z}$, for $\log(\ac{R})$,
$\log(\ac{z})$ and \ac{\phi}.  (Here the subscript `d' refers to the
donor model.)  After trimming their lower $1\%$, we compute the
average metallicity in bins in the $F_{d,\phi} \otimes F_{d,R} \otimes
F_{d,z}$ space of the donor.  We then use the actions of stars in
the target to construct its cumulative distribution functions of the
actions $F_{t,\phi}$, $F_{t,R}$ and $F_{t,z}$ (where now the subscript
`t' refers to the target model and we again use $\log$ of the radial
and vertical actions).  We then assign a metallicity to star particles
in the target by matching the cumulative distribution functions in the
donor and the target, ${\bf F_t} = {\bf F_d}$.

The resulting metallicity maps are shown in Fig. \ref{f:metalmaps}.
Both edge-on views bear strong resemblance to the metallicity maps of
the barred star-forming simulation of \citet{debattista+17} (their
Figure 26).  The metallicity distribution is more pinched than the
density in the side-on view.  Along the vertical axis the metallicity
has a declining profile.  In order to compare the metallicity gradient
with the MW's, the model needs to be rescaled because its bar is much
longer\footnote{We remind the reader that $N$-body simulations can
always be rescaled this way.  If length is rescaled as $L^\prime =
\alpha L$, then the time needs to be rescaled as $T^\prime =
\alpha^{3/2} T$.  In our case, with $\alpha = 5/9$, time is rescaled
by a factor of $0.4$.}.  Assuming the MW has a bar of $5~\kpc$
\citep{wegg+15}, we obtain a metallicity gradient of $-0.03$ dex/deg.,
about half the gradient measured by \citet{gonzalez+11}.  Considering
that this model is very different from the MW, this is a quite
promising result \citep[compared, for instance, with the much weaker
vertical gradients in the pure disc simulation
of][]{bekki_tsujimoto11} that implies that the actions can be
successfully used to assign metallicities to particles of B/P bulges
in $N$-body simulations.

The metallicity in the face-on view has hollows on the minor axis of
the bar, which are not observed in real galaxies.  In real galaxies
these regions are often found to have ``star formation deserts''
(SFDs) \citep{james_percival16, james_percival18} with ages older than
the average. Stellar populations in the SFDs are slowly refilled by
stars trapped by the bar at a later stage \citep{donohoe-keyes+19}.
Since our $N$-body model lacks significant late time bar growth, the
presence of these hollows in our model is perhaps not surprising. For
our fiducial model, the \feh\ hollow is only $\sim 0.1$ dex deep, and
may be relatively difficult to detect.


\section{Discussion} 
\label{s:discussion}

Our primary goals in this paper have been to map the initial actions
of an axisymmetric disc into the properties of the B/P bulge that
forms via the bar, and to develop a simple action-based method to
assign metallicity to disc particles in order to compare with
metallicity trends observed in B/P bulges in general, and the MW in
particular.  We have demonstrated a successful metallicity mapping
technique.  The study of the role of actions in the development of B/P
bulges leads to several insights.

\subsection{A statistical mechanical view}

The simplest result to acknowledge is just how reliable the {\it
initial} actions are at predicting the final distributions of
populations in the B/P bulge.  The bar and buckling instabilities are
violent processes (at least in these simulations), during which
angular momentum is redistributed, strong spirals propagate, and the
bar first grows exponentially in strength, then weakens abruptly.
During this time resonances must be sweeping the phase space as the
bar assembles and then weakens.  Orbital analyses of bars formed in
self-consistently evolving simulations typically find resonances well
populated \citep{valluri+16, abbott+17}.  In most of the simulations
presented here the disc potential dominates over the dark halo and the
gravitational perturbations must therefore be large.  Yet in spite of
all this roiling turmoil the structure that emerges in the B/P is
predetermined by the initial actions.  Since energy is being
transformed between in-plane and vertical motions, the actions cannot
be conserved.  We speculate therefore that the bar$+$buckling
instabilities drive the system towards a state of maximum entropy at
fixed angular momentum.  The clear dependence on the initial
conditions indicates that the dynamical instabilities are not violent
enough for complete mixing to occur. Testing this is beyond our scope
here, but is worth investigating.

And yet we should not permit ourselves to uncritically accept that the
bar$+$buckling instabilities generally occur in nature at the full
vigour of our pure $N$-body simulations.  It seems implausible that
galaxies often find themselves in a state so unstable that a
substantial reorganisation of their states begins immediately, and at
vigour.  Indeed simulations that include gas find bars forming more
slowly and buckling much more gently \citep{berentzen+98,
debattista+17}, even in a cosmological context \citep{debattista+19}.
The consequent energy redistribution that must occur is probably
milder.

Nonetheless, nature presents examples of buckling bars
\citep{erwin_debattista16, li+17}, including in the presence of gas.  
Whether these are events provoked by external perturbations, or driven
by purely internal secular evolution is as yet unclear.  However a
simple model of bar and B/P-bulge formation that assumes rapid violent
buckling is able to reproduce the observed fraction of B/P bulges and
buckling bars (the known number of which is still less than 10) amidst
barred galaxies \citep{erwin_debattista16}.  Complicating this simple
picture is the observation that the fraction of barred galaxies that
host B/P bulges rises rapidly at $\log(M_*/\Msun) \ga 10.4$, with no
evidence that the (present day) gas fraction matters
\citep{erwin_debattista17}.  Even more puzzling, this transitional
mass appears unchanged since redshift $z\simeq 1$
\citep{kruk+19}.  Clearly there is still much to learn about B/P bulge
formation and more detailed study of stellar populations in the MW,
and in external galaxies, may provide clues to a better understanding.

\subsection{Kinematic fractionation}

Building on the work of \citet{merritt_sellwood94},
\citet{debattista+17} argued that the vertical evolution of B/P bulges
is driven by the ability of stars to respond in phase with the
vertical forcing from a bending bar (through the buckling
instability).  Stars support a vertical perturbation of $m=2$ form so
long as they satisfy the condition:
\begin{equation}
\Omega_z > 2 (\Omega_\phi - \Omega_p),
\end{equation}
where $\Omega_z$ and $\Omega_\phi$ are the vertical and rotational
frequencies of the star, respectively, and $\Omega_p$ is the pattern
speed of the bar.  \citet{debattista+17} argued that $\Omega_z$
decreased much more slowly than $\Omega_\phi$ as the radial random
motion increases.  The separation of stellar populations by the bar
occurs because the larger asymmetric drift of stars with larger radial
random motions allows them to remain in tune with the vertical forcing
by the bar to greater heights. For this reason, they proposed that the
radial random motion of stars determined the vertical height stars can
reach during buckling, with radially hotter stars reaching larger
heights.

Recently, \citet{pdimatteo+19} argued that the vertical dispersion
before the bar forms is just as important as the radial dispersion.
We have shown here that the vertical thickening of populations is a
strongly monotonic function of \act{R}, but depends less strongly on
\act{z}.  Indeed we have seen that the populations with the largest
\act{z}\ generally become thinner after the formation of a B/P bulge,
and this population can be significant when the initial disc is thick.
A monotonically increasing vertical profile of \avg{\act{R}}\ develops
in most models as a result of B/P formation, while the vertical
gradient in \avg{\act{z}}\ is largely erased.  Both these facts argue
that the radial motions are much more important for the vertical
thickening of a population, as proposed by \citet{debattista+17}, and
at odds with the suggestion of \citet{pdimatteo+19}.

\subsection{Insights from bulge chemistry}

The vertical metallicity gradient \citep{zoccali+08, gonzalez+11,
johnson+11, johnson+13} and the metallicity dependence of the X-shape
\citep{ness+12, uttenthaler+12, rojas-arriagada+14} constrain 
the formation of the MW's bulge.  \citet{bekki_tsujimoto11} argued
that, in single disc models with a radial metallicity gradient, the
vertical mixing of stars caused by the bar produces too shallow a
vertical metallicity gradient. \citet{martinez-valpuesta_gerhard13}
subsequently showed that a steep radial metallicity gradient can
indeed be transformed into a vertical gradient similar to the
observed.  Liu et al. (in progress) reach a similar conclusion.  A
significant failure of this hypothesis is that the resulting bimodal
distance distribution in the bulge (the `X-shape') is more prominent
in metal-poor populations, whereas in the MW the X-shape is more
prominent in metal-rich populations \citep{ness+12, uttenthaler+12,
rojas-arriagada+14}.  An in-situ picture of the vertical gradient
relying on the superposition of a metal-rich thin disc and a
metal-poor thick disc \citep{bekki_tsujimoto11, pdimatteo16}, produces
a gradient via the transition from one population to another.
An alternative approach to producing the observed metallicity gradient
was presented by \citet{debattista+17} who found, in their simulation
with self-consistent star formation and chemistry, a vertical
metallicity gradient that resulted from the ability of older,
metal-poor stars to reach larger heights during the buckling
instability since they have larger radial random motion, compared to
slightly ($\sim 1~\Gyr$) younger stars.  The resulting X-shape is
better traced by the metal-rich populations, as observed.  They also
showed that as a consequence the metallicity distribution is more
peanut-shaped than the density itself \citep[see
also][]{athanassoula+17, buck+18, debattista+19}.  Such a metallicity
distribution has been confirmed in the edge-on galaxy NGC~4710, which
hosts a B/P bulge \citep{gonzalez+17}.

These trends help constrain models of the evolution of B/P bulges.  We
found that side-on maps of \act{z}\ and \act{R}\ are pinched and that
the bimodal populations at fixed heights are less separated with
increasing action. These are trends reminiscent of what is observed in
the MW's bulge.  If metallicity depends on \act{z}\ then a bimodality
forms but the formation of the bar and B/P bulge substantially weakens
the vertical gradient \citep{bekki_tsujimoto11}.  In this case we find
that the initial vertical gradient is substantially weakened.  Instead
metallicity based on \act{R}\ will produce a strong and monotonic
vertical gradient, as observed in the MW.  This is an efficient way to
generate vertical gradients from a system in which the in-plane random
motions are a function of age by the time the bar forms.  However we
argue that \act{R}\ by itself is not sufficient to properly assign
metallicities to particles.  We show that assigning metallicities
using all three actions results in a model that has the right X-shape
metallicity variation and a vertical metallicity gradient not very
different from the MW's.

\subsection{Outlook for action-based metallicity assignment}

At this time, pure $N$-body (collisionless) simulations with $10^8$
particles do not represent significant computational difficulties
\citep[e.g.][]{sellwood12, donghia+13}.  The collisionless cosmological 
simulation of \citet{potter+17} employed $2 \times 10^{12}$ particles,
which required only a modest 350,000 node hours to complete.  On the
other hand, hydrodynamical simulations with this number of particles
constitute a huge computational effort.  For instance, the Illustris
TNG100 simulation, one of the IllustrisTNG flagship runs with $1.2
\times 10^{10}$ resolution elements, required $\sim 18$M cpu hours to
complete \citep{dnelson+18, marinacci+18, pillepich+18, naiman+18,
springel+18}, with other runs in the IllustrisTNG suite requiring
longer for varying resolutions.  While these are extreme examples (and
also differ in their resolutions), they demonstrate the very different
cost of running pure $N$-body versus hydrodynamical simulations.
Action-based stellar-population modelling therefore holds great
promise for helping to understand the distribution of stellar
populations within B/P bulges by permitting rapid studies in parameter
space using $N$-body simulations.

We have presented a simple experiment in assigning chemistry using the
three actions.  The model exhibits many of the trends of real B/P
bulges.  Since our simulations are pure $N$-body, with no star
formation, their evolution covers only stellar populations extant at
the time of bar formation.  Because the bar grows by trapping disc
particles, stars younger than the bar are not represented in the
simulations.  However, since stars forming after the bar do not
substantially participate in the B/P bulge \citep{debattista+17}, this
omission does not seriously affect our results on the stellar
populations of the B/P bulge itself.  On the other hand, the stellar
populations of the disc, where star formation continues even after the
bar forms, cannot reliably be captured by our action-based metallicity
assignment.  Nevertheless, we conclude that within the main bulk of
the B/P bulge itself, the action-based metallicity assignment we have
developed is a very useful tool for studying the stellar distribution.
The method can be adapted in a variety of ways, including replacing
the donor with an analytic prescription for the metallicity-action
relation, including more moments in the metallicity distribution
function of the donor, or mapping based on machine learning
techniques.

\subsection{Summary}

We have computed the actions of star particles in the initial
conditions of a suite of pure $N$-body simulations.  We followed the
particles to the point where a B/P bulge forms, and examined how the
initial actions were distributed in the final B/P bulge.  We also used
a star-forming simulation, matched in action space, to assign
metallicity to the particles of a pure $N$-body simulation.  Our main
results are:
\begin{itemize}
\item Both the radial action, \act{R}, and the vertical action, \act{z}, 
separate stellar populations such that those with lower actions
support a stronger, more elongated bar while those with higher
actions host a weaker bar.  The bar strength has a greater dynamic
range when stellar populations are separated by \act{R}\ than by
\act{z} (see Section \ref{ss:barstrength}).
\item The central vertical profile of \avg{\act{z}}\ is substantially 
flattened by the formation of the B/P bulge.  A vertical gradient
develops in \avg{\act{\phi}}\ (and also in \avg{R_0}, the mean radius
of stars at $t=0$), which reaches a peak and then declines at larger
heights.  The initially rather flat profile of \avg{\act{R}}\ is
transformed into a monotonically rising profile by the B/P bulge
formation (see Sections \ref{ss:actionmaps} and \ref{ss:gradients}).
\item Stellar populations separated by either \act{R}\ or \act{z}\ display 
an increasing separation of the peaks of the bimodality with
increasing height.  The separation is stronger for the populations
with lower \act{R}\ and lower \act{z}, and higher \act{\phi}.  This
different trend with \act{\phi}\ would imply a greater separation in
the X-shape in metal-poor populations if metallicity depended on
angular momentum, or radius (see Section \ref{ss:bimodalities}).
\item The thickness at the end of the simulation, and the overall
vertical thickening, increases monotonically with \act{R}.  This is
the characteristic of kinematic fractionation described by
\citet{debattista+17}.  In contrast, the thickening is peaked at
intermediate \act{z}, decreasing to larger values and becoming
negative (\ie\ the populations become thinner) for the highest
\act{z}.  Conversely, the populations with the largest \act{R}\
radially cool (see Section \ref{ss:heating}).
\item Only models with a narrow range of \act{R}\ diverge from these trends, 
and exhibiting a weak dependence of the bar strength and X-shape on
the actions.  These contrary behaviours serve to further illustrate
the importance of the radial action in the development of the trends
observed in stellar populations observed in the MW (see Section
\ref{s:simsuite}).
\item We demonstrate a simple action-based mapping for setting the  
metallicity of star particles in a pure $N$-body simulation based on
their actions using the metallicity of particles in a star-forming
simulation.  The resulting metallicity map of the galaxy seen edge-on
with the bar seen side-on is pinched, matching observations in real
galaxies and hydrodynamical models of barred galaxies (see Section
\ref{ss:tagging}).
\end{itemize}

In future papers we will present further analysis of these models and
employ the action-based metallicity tagging to develop models of the
MW's bulge chemistry.


\bigskip
\noindent
{\bf Acknowledgements.}

\noindent
V.P.D. and L.B.S. are supported by Science and Technology Facilities
Council Consolidated grant \#~ST/R000786/1.  D.J.L. was supported for
part of this project by a UCLan UURIP internship.  The simulations was
run at the DiRAC Shared Memory Processing system at the University of
Cambridge, operated by the COSMOS Project at the Department of Applied
Mathematics and Theoretical Physics on behalf of the STFC DiRAC HPC
Facility (www.dirac.ac.uk). This equipment was funded by BIS National
E-infrastructure capital grant ST/J005673/1, STFC capital grant
ST/H008586/1 and STFC DiRAC Operations grant ST/K00333X/1. DiRAC is
part of the National E-Infrastructure.  We thank Larry Widrow and
Nathan Deg for their assistance with {\sc galactics}.

\bigskip 
\noindent

\bibliographystyle{aj}
\bibliography{ms.bbl}

\label{lastpage}

\end{document}